\documentclass[aps,prl,twocolumn,superscriptaddress,groupedaddress]{revtex4-2}
\usepackage{graphicx}  % needed for figures
\usepackage{mwe}
\usepackage{bm}        % for math
\usepackage{amssymb}   % for math
\usepackage{physics}
\usepackage{float}
\usepackage{verbatim}
\usepackage[dvipsnames]{xcolor}
\usepackage[mathscr]{euscript}
\usepackage{array}

\newcommand{\PreserveBackslash}[1]{\let\temp=\\#1\let\\=\temp}
\newcolumntype{C}[1]{>{\PreserveBackslash\centering}p{#1}}
\newcolumntype{R}[1]{>{\PreserveBackslash\raggedleft}p{#1}}
\newcolumntype{L}[1]{>{\PreserveBackslash\raggedright}p{#1}}

\begin{document}

\title{Cryptographic approach to Quantum Metrology}

\author{Nathan Shettell$^{1}$, Elham Kashefi$^{1,2}$ and Damian Markham$^{1,3}$}
\affiliation{$^1$LIP6, CNRS, Sorbonne Université, 4 place Jussieu, 75005 Paris, France}
\affiliation{$^2$School of Informatics, University of Edinburgh, EH8 9AB Edinburgh, United Kingdom}
\affiliation{$^3$Japanese-French Laboratory for Informatics, CNRS, National Institute for Informatics, University of Tokyo, Tokyo, Japan}

\date{\today}

\begin{abstract}
We consider a cryptographically motivated framework for quantum metrology in the presence of a malicious adversary. We begin by devising an estimation strategy for a (potentially) altered resource (due to a malicious adversary) and quantify the amount of bias and the loss in precision as a function of the introduced uncertainty in the resource. By incorporating an appropriate cryptographic protocol, the uncertainty in the resource can be bounded with respect to the soundness of the cryptographic protocol. Thus the effectiveness of the quantum metrology problem can be directly related to the effectiveness of the cryptography protocol. As an example, we consider a quantum metrology problem in which resources are exchanged through an unsecured quantum channel. We then construct two protocols for this task which offer a trade-off between difficulty of implementation and efficiency.
\end{abstract}

%\pacs{}
\maketitle

By using quantum systems as probes to measure unknown parameters, quantum metrology offers precision and sensitivity not possible classically \cite{giovannetti2006,giovannetti2011}. On the other hand, quantum cryptography uses quantum systems to detect, and hence avoid the effects of, malicious behaviour \cite{pirandola2020advances}. 
It is natural to want to combine these two features, in particular with the advance of a quantum internet \cite{wehner2018quantum} where we expect entangled states to act as resources for a variety of tasks, including networks of sensors, and to be shared around over potentially unsecured channels. There have been exciting steps in this direction (e.g. \cite{komar2014quantum,huang2019,xie2018,takeuchi2019,okane2020,yin2020}); however, to date there is a lack of clear framework and definitions of security, which, as we will see, poses the danger that protocols are left open to unforeseen attacks.

In a (potentially) malicious setting, the quantum system used as a probe is vulnerable to attacks which alter said probe in an undesirable and unknown fashion. This added uncertainty directly affects the precision related to the estimation portion of the quantum metrology problem. However, quantifying the effects of said uncertainty is far from trivial. In theory, the quantum Fisher information (QFI) is an ultimate bound on the attainable precision when estimating the unknown parameter \cite{braunstein1994, luo2000}. In practice however, the precision is related to the estimation strategy (the methodology of treating the measurement results). The bound set by the QFI can only be obtained when the estimation strategy is unbiased, that is, the expected estimate is the true value of the unknown parameter. It is not immediately evident how to process the measurement results if the quantum system had been tampered with in an unknown fashion. More so, how does one ensure their estimator is unbiased? This task is more difficult than a noisy quantum metrology problem \cite{escher2011a, demkowicz2012}, where noise alters the quantum state according to a specified model. We are proposing a setting where the altercation by an adversary is completely unknown and malicious.

In this article we formalize \textit{cryptographically secure quantum metrology}. We incorporate core concepts frequently used in quantum cryptography within this framework, including (i) \emph{privacy}: the notion that an adversary cannot gain information, (ii) \emph{soundness}: a measure of detecting any malicious activity, and (iii) \emph{integrity}: the ability to retain the quantum state and functionality in the presence of malicious adversaries. Specifically, we propose an estimation strategy when the added uncertainty is small, which can be certified incorporating an appropriate cryptographic protocol. From a quantum metrology perspective, we show that the integrity of the estimation strategy can be related to the notion of soundness.

To gauge the utility of the presented framework, we present two cryptographic protocols for a quantum metrology problem in which resources are shared over unsecured quantum channels. We imagine this setting arising in quantum networks where different nodes have asymmetric capabilities, for example one node may be optimized for initializing quantum states, where another node is designed for quantum measurements. From a functional standpoint, the two protocols are nearly identical; however, the encryption and decryption methods vary drastically from a complexity standpoint and ease of implementation. We show that both protocols are completely private and derive a bound on the soundness of the protocols. Combining everything, we derive a bound on the required number of additional resources to (approximately) recover the same precision one would obtain in a setting sans adversary. We show that the more complex protocol requires an additional log-linear number of total qubits to maintain the same level of precision, whereas the simpler protocol requires an additional quadratic number of total qubits to achieve the same feat. Furthermore, this setting and the protocols naturally extend to more involved settings over a multipartite quantum network, a popular topic in the quantum metrology and sensing community \cite{komar2014quantum, ge2018, proctor2018, rubio2020}.

%\section{Parameter Estimation}

\begin{figure}
\includegraphics[width=0.43\textwidth]{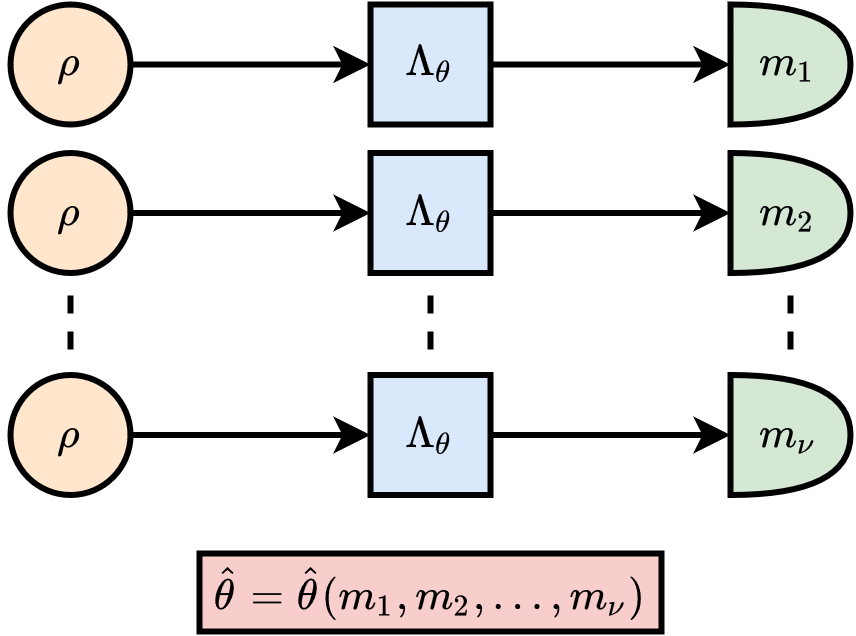}
\caption{An unknown parameter $\theta$ is encoded into a quantum state $\rho$ via a CPTP map $\Lambda_\theta$, after which a measurement is performed. This is repeated $\nu$ times and the measurement results $m_1,m_2,\ldots,m_\nu$ are used to construct an estimate $\hat{\theta}$.}
\label{fig:QuantumMetrology}
\end{figure}

We begin by reviewing a standard estimation strategy of inferring the value of an unknown parameter using an observable. This will be the foundation of the proposed estimation strategy in the presence of an adversary. This strategy is commonly referenced in phase estimation problems as the precision limited by the QFI can be saturated with a simple measurement scheme \cite{giovannetti2006, toth2014}. For all intents and purposes, we will remain completely general, i.e., an unknown parameter is encoded into an $n$ qubit quantum state $\rho$ via a completely positive trace-preserving (CPTP) map $\Lambda_\theta: \rho \rightarrow \rho_\theta$. Next, $\rho_\theta$ is measured with respect to the eigenbasis of an observable $O$. Assuming that the measurement is appropriately chosen, the distribution of the measurement results will be dependent on $\theta$. By repeating the prepare and measure portion of the quantum metrology problem sufficiently many times, one can construct a high precision estimate $\hat{\theta}$. The process is illustrated in Figure~\ref{fig:QuantumMetrology}. We use the notation $\hat{\square}$ to indicate an estimate and to differentiate it from the true value.

Specifically, one uses the measurement results to estimate the quantity $f(\theta)=\expval{O}_{\rho_\theta} = \Tr ( O \rho_\theta )$, after which the inverse function $f^{-1}$ is applied to obtain an estimate of $\theta$ \cite{toth2014}. We denote the initial estimate by $\hat{f}$, thus $\hat{\theta}=f^{-1}(\hat{f})$. Suppose that $O$ has eigenvalues $\{ o_i \}$ with associated projectors of its eigenbasis $\{ \Pi_i \}$. If the $j$th measurement outcome is $\Pi_i$ one sets $m_j=o_i$, from which one constructs the estimate
\begin{equation}
    \label{eqn:obsestimate}
    \hat{f}=\frac{1}{\nu} \sum_{j=1}^{\nu} m_j.
\end{equation}
The above estimate is unbiased because $m_j=o_i$ occurs with probability $\Tr \big( \Pi_i \rho_\theta \big)$, and hence $\mathbb{E}(m_j)=f(\theta) \; \forall j$. With sufficient measurement results, $\nu \gg 1$, $\hat{f}$ will fluctuate close to $f(\theta)$, and the error propagation formula states that the mean squared error with respect to the estimation of $\theta$ is
\begin{equation}
\Delta^2 \hat{\theta} = \mathbb{E} \big( (\hat{\theta}-\theta)^2 \big) = \frac{\Delta^2\hat{f}}{ |\partial \expval{O}_{\rho_\theta}|^2} =\frac{\Delta^2 O_{\rho_\theta}}{\nu |\partial \expval{O}_{\rho_\theta}|^2},
\label{eqn:Acc}
\end{equation}
where $\Delta^2 O_{\rho_\theta}=\expval*{O^2}_{\rho_\theta} - \expval*{O}_{\rho_\theta}^2$ is the variance of the observable $O$ with respect to $\rho_\theta$.

Next, consider an analogous setting where a malicious adversary may have tampered with the resources, such that the measurement results are obtained from the $n\nu$ qubit quantum state $\rho^\prime$ (as opposed to $\rho_\theta^{\otimes \nu}$ in the ideal setting). Notice that we do not assign a subscript of $\theta$ to this resource, or assume many copies of identical copies using the superscript $\otimes \nu$. This is because no assumptions are made about the actions of the adversary, e.g., they may replace $\rho_\theta$ with a completely unknown state, nor do they necessarily act uniformly, and finally, they may replace the overall resource by a giant entangled quantum state. Of course, for the estimate to be meaningful in the presence of an adversary, we impose that the quantum state is close to the ideal resource,
\begin{equation}
    \label{eqn:bound}
    \frac{1}{\nu}\sum_{j=1}^\nu \mathscr{D}(\rho^{\prime (j)},\rho_\theta) \leq \varepsilon,
\end{equation}
where $\rho^{\prime (j)}$ is a reduced state with all but the $j$th block of $n$ qubits of $\rho^\prime$ traced out, and $\mathscr{D}$ is the trace distance. We will later show that this can be certified using an appropriate cryptographic protocol. We use the notation $\square^\prime$ to indicate a quantity in the malicious setting.

As eluded to earlier, devising an unbiased estimation strategy in the presence of an adversary is far from trivial. Instead, we propose the same strategy one would use in the ideal setting, i.e., constructing an estimate $\hat{f}^\prime$ of $f(\theta)$ and setting $\hat{\theta^\prime}=f^{-1}(\hat{f}^\prime)$. Here, $\hat{f}^\prime$ is constructed in the same way of $\hat{f}$: the $j$th block of $n$ qubits of $\rho^\prime$ is measured with respect to the eigenbasis of $O$ and if $\Pi_i$ is the outcome of the measurement, then $m_j^\prime=o_i$ and $\hat{f}^\prime=\frac{1}{\nu}\sum_{j=1}^\nu m_j^\prime$.

Such an estimate is (potentially) biased because $\mathbb{E}(m_j^\prime)=\Tr(O \rho^{\prime (j)})$ is not guaranteed to be equal to $f(\theta)$ for all $j$. Eq.~\eqref{eqn:bound} imposes that $\rho^\prime$ is close to $\rho_\theta^{\otimes \nu}$, and although there may be a better estimation strategy for a specific resource which satisfies Eq.~\eqref{eqn:bound}, due to the lack of additional information, the best strategy is to construct the estimate as if it was the ideal resource. Nonetheless, the amount of bias can be bounded using the fact that for any POVM $\{F_i \}$ and two quantum states $\rho_1$ and $\rho_2$ \cite{NC}
\begin{equation}
    \frac{1}{2}\sum_i |\Tr \big( F_i(\rho_1-\rho_2) \big)| \leq \mathscr{D}(\rho_1,\rho_2),
\end{equation}
and thus
\begin{equation}
\label{eq:biasderivation}
\begin{split}
    |\mathbb{E}(\hat{f}^\prime - \hat{f})| &= \frac{1}{\nu} \big| \sum_{j=1}^\nu  \Tr \big(O(\rho^{\prime (j)}-\rho_\theta) \big) \big| \\
    & \leq \frac{2o}{\nu} \sum_{j=1}^\nu \mathscr{D}(\rho^{\prime (j)},\rho_\theta) \\
    &\leq 2o \varepsilon,
\end{split}
\end{equation}
where $o$ is the maximum magnitude of the eigenvalues of $O$.

Assuming that $\varepsilon^2 \ll 1$ is sufficiently small, $\hat{f}^\prime$ will still fluctuate close enough to $f(\theta)$ such that the linear expansion of $f^{-1}(\hat{f}^\prime)$ is a valid approximation. This has two implications: the first is that the bias introduced to the estimation of $\theta$ is linear,
\begin{equation}
    |\mathbb{E}(\hat{\theta}^\prime - \hat{\theta})|=\frac{|\mathbb{E}(\hat{f}^\prime - \hat{f})|}{|\partial \expval{O}_{\rho_\theta}|},
\end{equation}
and the second is that the precision is similarly obtained from the error propagation formula,
\begin{equation}
    \Delta^2 \hat{\theta}^\prime=\frac{\Delta^2\hat{f}^\prime}{ |\partial \expval{O}_{\rho_\theta}|^2}.
\end{equation}
The difference in precision can be computed using a similar technique as per the bound on the bias. Note that
\begin{equation}
\begin{split}
\Delta^2 \hat{f}^\prime &= \mathbb{E} \big( (\hat{f}^\prime -f(\theta))^2 \big) \\
&= \mathbb{E} \big( (\hat{f}^\prime -\mathbb{E}(\hat{f}^\prime)+\mathbb{E}(\hat{f}^\prime)-\mathbb{E}(\hat{f}))^2 \big) \\
&= \mathbb{E} ( \hat{f}^{\prime 2} ) - \mathbb{E} ( \hat{f}^\prime )^2 +  \mathbb{E} ( \hat{f}^\prime - \hat{f}) ^2 \\
&= \frac{1}{\nu^2}\sum_{j=1}^\nu \Tr \big( \tilde{O} \rho^{\prime (j)} \otimes \rho^{\prime (j)}  \big) +  \mathbb{E} ( \hat{f}^\prime - \hat{f}) ^2,
\end{split}
\end{equation}
where $\tilde{O}= O^2 \otimes \mathbb{I} - O\otimes O$ is an observable whose eigenvalues have a magnitude bounded by $2o^2$. Using the same reasoning as found in Eq.~\eqref{eq:biasderivation}, it follows that
\begin{equation}
\begin{split}
    | \Delta^2 \hat{f}^\prime - \Delta^2 \hat{f} \big|
    \leq &  \frac{4o^2}{\nu^2}\sum_{j=1}^\nu \mathscr{D} (\rho^{\prime(j)} \otimes \rho^{\prime(j)}, \rho_\theta \otimes \rho_\theta) + 4o^2 \varepsilon^2 \\
    \leq & 8o^2 \varepsilon \nu^{-1} + 4o^2 \varepsilon^2,
\end{split}
\end{equation}
where the triangle inequality $\mathscr{D}(\rho_1 \otimes \rho_1,\rho_2 \otimes \rho_2) \leq 2 \mathscr{D}(\rho_1,\rho_2)$ is used in the final inequality.

\textit{Theorem 1: Bounds on the bias and difference in precision.} If a resource satisfies Eq.~(\ref{eqn:bound}), the estimate bias is bounded via
\begin{equation}
\big| \mathbb{E}(\hat{\theta}^\prime) - \mathbb{E}(\hat{\theta}) \big|  \leq \frac{2o\varepsilon}{|\partial \expval{O}_{\rho_\theta}|}
\label{eqn:AccBias}
\end{equation}
and the difference in precision is bounded via
\begin{equation}
\big| \Delta^2\hat{\theta}^\prime-\Delta^2\hat{\theta} \big| \leq  \frac{4o^2 (2\varepsilon \nu^{-1}+\varepsilon^2 )}{|\partial \expval{O}_{\rho_\theta}|^2}.
\end{equation}

We immediately observe that the added biases do not vanish as $\nu$ increases. This is due to the fact that $\varepsilon$ is not explicitly dependent on $\nu$, and thus uncertainty can be introduced to each measurement outcome. Additionally, the biases scale inversely with $|\partial \expval{O}_{\rho_\theta}|$; this is because any uncertainty in the estimate of $f(\theta)$ translates to significantly more uncertainty after inverting the estimate when $\expval{O}_{\rho_\theta}$ is close to a local extrema. The final observation we make is that if $\varepsilon \leq \sqrt{\nu^{-1}}$, then a similar level of precision is achieved in an adversarial setting. This result is not surprising because $\Delta^2 \hat{\theta} \rightarrow 0$ as $\nu \rightarrow \infty$. Therefore, to maintain the desired precision, one requires that the uncertainty in the overall resource be infinitely small: $\varepsilon \rightarrow 0$. Note that \textit{Theorem~1} is not only of interest for malicious sources, but it holds for any source which satisfies Eq.~(\ref{eqn:bound}). For example, this may be simpler to utilize than devising an unbiased estimator for a noisy system with a complex noise model.

In the remainder of this article, we motivate the introduced framework for a parameter estimation strategy in the presence of a malicious adversary. We consider a quantum metrology problem where the resources are exchanged over an unsecured quantum channel, the necessity for which may be because of asymmetric hardware capabilities. To circumvent a malicious eavesdropper from biasing the quantum state passing through the quantum channel, we propose two protocols which prevent the eavesdropper from obtaining information about the parameter (\textit{privacy}) and allows the trusted parties to detect any alterations done by the eavesdropper with high probability (\textit{soundness}). As a result, we will show that we can certify that the effective resource used is close to the ideal resource, as per Eq.~\eqref{eqn:bound} (\textit{integrity}).

The construction of the two protocols makes use of quantum authentication schemes \cite{barnum2002,broadbent2016,gheorghiu2019}. The two protocols we outline are (i) a modified version of the trap code \cite{broadbent2013}, and (ii) a modified version of the Clifford code \cite{aharonov2017}. Both protocols have a simplistic algebraic representation, while having substantially different requirements for implementation. The Clifford code uses an arbitrary operator from the Clifford group, $\mathcal{C}_m$, to encode an $m$ qubit quantum state. In contrast, our version of the trap code uses an arbitrary operator from the group of single qubit Cliffords, $\mathcal{C}_1^{\otimes m}$. As expected, the Clifford code leads to much stronger soundness and integrity statements due to the additional entanglement gained from the encryption, which conversely makes implementation much more difficult.

\begin{figure}
\includegraphics[width=0.48\textwidth]{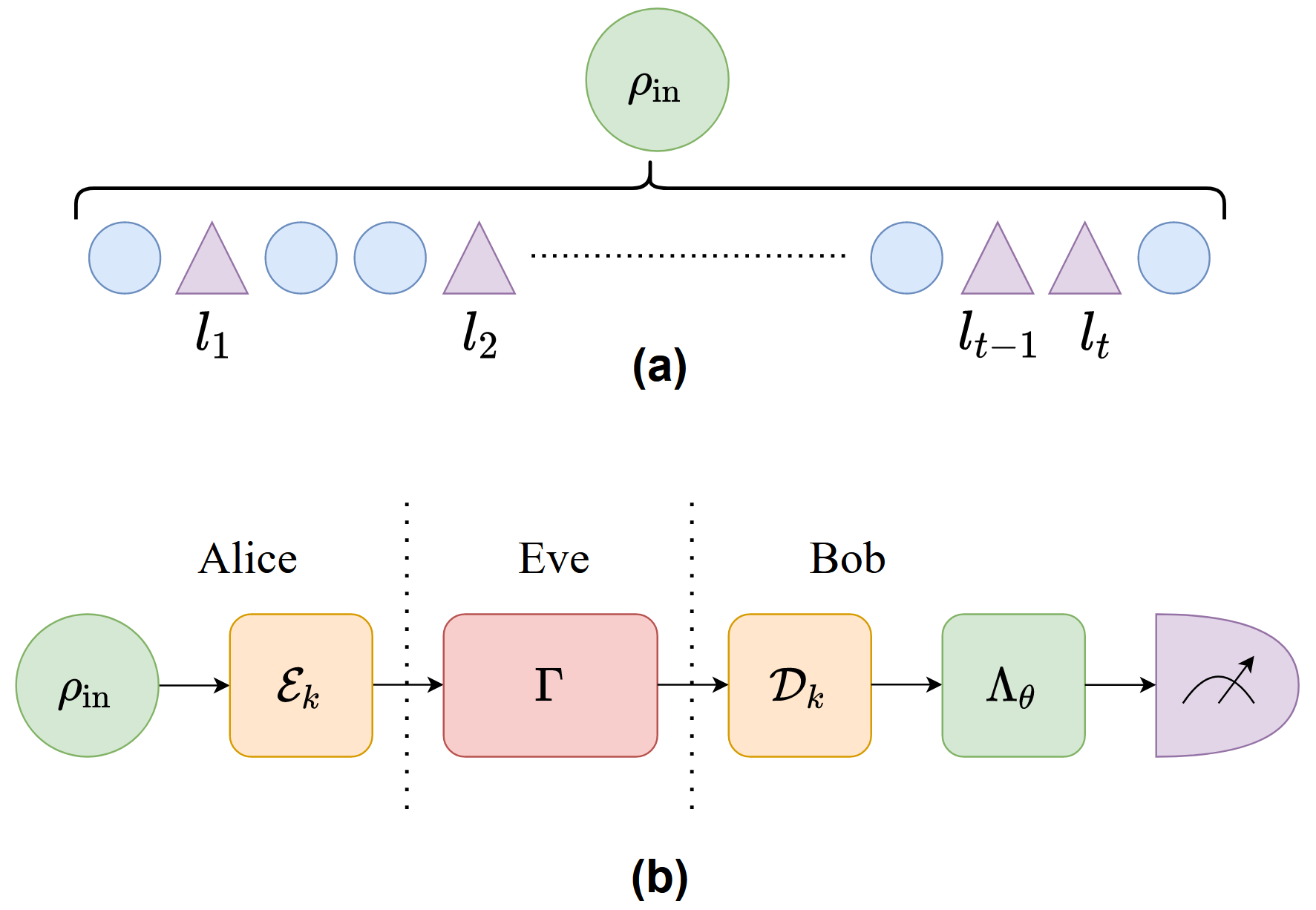}
\caption{(a) The quantum state that Alice initializes, $\rho_\text{in}$, is a combination of $t$ ancillary flag qubits (randomly positioned) as well as the quantum state $\rho$ intended for quantum metrology. The flag qubits are indexed at positions $l_1,l_2,\ldots,l_t$ and give a means of detecting malicious interference. (b) Alice and Bob share a classical key $k$ which corresponds to the encryption ($\mathcal{E}_k$) and decryption ($\mathcal{D}_k$) operations. These operations prevent a malicious eavesdropper, Eve, from accessing any information about the quantum state being passed through the quantum channel, despite Eve being able to perform any CPTP map $\Gamma$. After performing the decryption operation, Bob encodes $\Gamma_\theta$ onto the appropriate qubits and performs the relevant measurements. If the measurement result of any of the flag qubits is an unexpected output, then a malicious adversary must have tampered with the quantum channel.}
\label{fig:QASMetrology}
\end{figure}

For both protocols, Alice prepares an input state $\rho_\text{in}$, which is a combination of the quantum state designated for quantum metrology, as well as $t$ ancillary flag qubits. An example of an input state is depicted in Figure~\ref{fig:QASMetrology}(a). The flag qubits are set to the $\ket{0}$ state and are intended to act as traps. Their deterministic measurement results are used to determine with high probability whether or not a malicious eavesdropper tampered with the quantum channel.

After creating the input state, Alice encrypts it with a Clifford operation. The set from which the Clifford operation is chosen is dependent on the protocol. By encrypting the quantum state before using the quantum channel, it prevents any malicious eavesdroppers from extracting information with regard to the quantum state. Bob can recover the original quantum state upon receipt by performing the decryption operation, which in this case is the inverse of the Clifford that Alice applied. Finally, Bob measures the ancillary flag qubits in the computation basis. Bob will utilize the remaining quantum state for quantum metrology (parameter encoding, measuring an observable and constructing an estimate) if all of the ancillary flag qubits measure as $\ket{0}$; otherwise they discard the quantum state as someone must have tampered with the quantum channel. This process is illustrated in Figure~\ref{fig:QASMetrology}(b).

\begin{center}
    \textit{Summary of the Protocols:}
\end{center}

\begin{enumerate}

\item Prior to implementing the protocol, Alice and Bob randomly select a key $k \in \mathcal{K}$, which is linked to an encryption operator $\mathcal{E}_{k}$. Additionally, for the trap code, the key contains information about a tuple $\vec{l}=(l_1,\ldots,l_t)$ of length $t$; this tuple contains the index locations of the ancillary flag qubits.

\begin{enumerate}
    \item For the trap code, $\mathcal{E}_k \in \mathcal{C}_1^{\otimes m}$.
    \item For the Clifford code, $\mathcal{E}_k \in \mathcal{C}_m$.
  \end{enumerate}

\item Alice creates the $m=n+t$ qubit state $\rho_\text{in}$ by inserting $t$ ancillary flag qubits $\ket{0}$ at the positions indexed by $\vec{l}$, and the remaining $n$ qubit state $\rho$ is the quantum state designated for quantum metrology.

\begin{enumerate}
    \item When implementing the trap code, it is important that $\vec{l}$ is randomly chosen because the encryption operation does not generate entanglement.
    \item When implementing the Clifford code, one can set $\vec{l}$ to be fixed. This is because the encryption will generate entanglement between the ancillary qubits and the rest of the quantum state.
  \end{enumerate}

\item Alice encrypts the input state by applying the Clifford operator $\mathcal{E}_k$. Bob decrypts the quantum state by applying the inverse operator $\mathcal{E}_k^\dagger$ upon receipt.

\item Bob measures the ancillary flag qubits in the computational basis. The result is accepted if $\dyad{0}^{\otimes t}$ is measured. The quantum state is discarded otherwise.

\item If the result is accepted, Bob encodes the unknown parameter in the remaining qubits, which are measured in an appropriate basis to ultimately construct an estimate of said unknown parameter after sufficiently many copies have been measured.

\end{enumerate}

\textit{Theorem 2: Privacy.} Let $\rho_E$ be the $m$ qubit quantum state accessible to Eve; then, at any point Eve cannot extract any information because
\begin{equation}
\mathbb{E} \big( \rho_E \big)=\frac{\mathbb{I}}{2^m}.
\end{equation}
The proof of the result is given in the Appendix A. Note that Figure~\ref{fig:QASMetrology}(b) indicates that Bob performs $\Lambda_\theta$ after Alice sends $\rho$ through the quantum channel. However, we can equally consider Alice sending $\rho_\theta$ through the quantum channel. There the same protocol(s) could be used, and having complete privacy is integral to prevent Eve from learning any information about $\theta$.

The other desired characteristic of the protocols is soundness: the ability to detect any alterations by a malicious eavesdropper with high probability. We say the protocol has soundness $\delta$ if, for any malicious attack $\Gamma$,
\begin{equation}
\label{eqn:integrity}
\frac{1}{|\mathcal{K}|} \sum_{k \in \mathcal{K}} \Tr \big( \Pi_k \rho_\text{out}(k,\Gamma) \big) \leq \delta,
\end{equation}
where $\mathcal{K}$ is the set of all possible classical keys $k$, $\rho_\text{out}(k,\Gamma)$ is the outputted quantum state after undergoing attack $\Gamma$ and encryption described by key $k$, and $\Pi_k$ is the projector onto the accepted output of the ancillary flag qubits and the orthogonal complement of the ideal output. In Appendix B we show that the trap code has soundness $\delta_{\text{trap}}=\frac{3n}{2t}$, and the Clifford code has soundness $\delta_{\text{cliff}}=\frac{1}{2^t}$. These results are derived using twirling lemmas \cite{dankert2009} and a general Kraus decomposition of $\Gamma$.

It is important to understand that soundness is not equivalent to the infidelity between the output and the ideal output. Without loss of generality if the ideal output is a pure state, we can re-write Eq.~(\ref{eqn:integrity}) as
\begin{equation}
\mathbb{E} \big[ p_\text{acc} \cdot \big( 1-F(\rho_\text{id},\rho^\prime) \big) \big] \leq \delta,
\end{equation}
where $\rho$ is the ideal output, $p_\text{acc}$ is the probability of accepting, and $\rho^\prime$ is the output quantum state conditional on the ancillary flag qubits measurement resulting in accept, where the latter two quantities are dependent on the key $k$, but the subscripts are dropped for clarity. One may desire a stricter bound on the fidelity $F(\rho_\text{id},\rho^\prime)$ that is conditional on measuring the ancillary flag qubits in an accepted state. However, this is impossible with no restrictions on $\Gamma$ \cite{note1}. Nevertheless, we make the following claim:
\begin{equation}
\label{eqn:integrity2}
p_\text{acc} > \alpha \; \; \forall k \hspace{5pt} \Rightarrow \hspace{5pt} 1-F(\rho_\text{id},\rho^\prime)\leq \delta/\alpha,
\end{equation}
where we have relabeled $\rho^\prime$ to be the expected output state if the ancillary measurement results in accept. This necessary step is frequently used by the verification community \cite{zhu2019}, and $\alpha$ is sometimes referred to as the statistical significance.

Recall that the quantum metrology protocol requires $\nu$ copies of $\rho_\theta=\Lambda_\theta ( \rho_\text{id})$. Because the trace distance is contractive under CPTP maps, by utilizing inequalities between fidelity and trace distance \cite{fuchs1999}, we can re-write Eq.~(\ref{eqn:integrity2}) as
\begin{equation}
\label{eqn:newbound}
\frac{1}{\nu}\sum_{j=1}^\nu \mathscr{D}(\rho^{\prime (j)},\rho_\theta) \leq \sqrt{\frac{\delta}{\alpha}},
\end{equation}
where the superscript $(j)$ denotes the expected quantum state after the $j$th output of the protocol is encoded by Bob. By combining the above with Theorem 1, we obtain the following:

\textit{Theorem 3: Integrity of the quantum metrology problem using the described cryptography protocols.} After implementing the cryptographic protocols described earlier, the quantum metrology scheme will have a bias bounded by
\begin{equation}
\big| \mathbb{E}(\hat{\theta}^\prime) - \mathbb{E}(\hat{\theta}) \big|  \leq \frac{2o}{|\partial \expval{O}_{\rho_\theta}|} \sqrt{\frac{\delta}{\alpha}},
\end{equation}
and the difference in precision (compared to the ideal setting) is bounded via
\begin{equation}
\big| \Delta^2\hat{\theta}^\prime-\Delta^2\hat{\theta} \big| \leq  \frac{4o^2}{|\partial \expval{O}_{\rho_\theta}|^2} \big( 2\sqrt{\frac{\delta}{\alpha}} \nu^{-1}+\frac{\delta}{\alpha} \big),
\label{eqn:PrecBias}
\end{equation}
with $\delta_\text{trap}=\frac{3n}{2t}$ when using the trap code and $\delta_\text{cliff}=\frac{1}{2^t}$ when using the Clifford code. Thus, the number of ancillary qubits to retain the same level of precision is $t_\text{trap} \geq \frac{3n \nu}{2\alpha}$ using the trap code, and $t_\text{cliff} \geq \log_2 \frac{\nu}{\alpha}$. As the protocol is repeated $\nu$ times, this translates to a quadratic increase in qubits for the trap code and a log-linear increase for the Clifford code.

The work in \cite{huang2019} also addresses the distribution of entangled resources over quantum channels for quantum metrology; however, with a more restricted Bob, so that the measurement is also left to Alice, requiring the state be sent back to Alice once Bob has done the encoding. This could be desirable if we require that Bob does get the information about the parameter. Though shown to be secure against particular attacks considered, the protocol in \cite{huang2019} is unfortunately insecure in  general. In particular, in Appendix D, we show that there exists an attack which is undetectable and prevents the trusted parties from learning any information about the unknown parameter, while the eavesdropper can learn some information about the unknown parameter. This highlights the need for the approach in this work. Our protocol is easily extended to cover this scenario and we show in Appendix C that the soundness of the protocols in this situation is $\delta_\text{trap}=\frac{9n}{4t}$ for the trap code and $\delta_\text{cliff}=\frac{1}{2^t}$ for the Clifford code.

\begin{figure}
\includegraphics[width=0.3\textwidth]{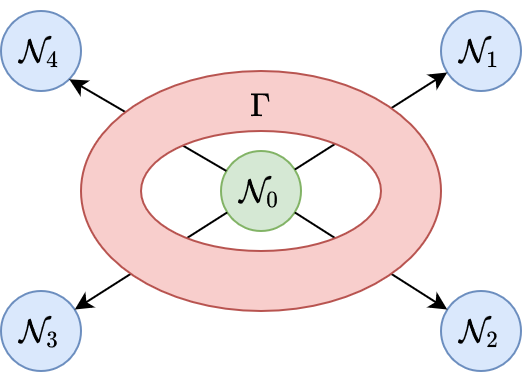}
\caption{Generalization to a multipartite setting, where a central node $\mathcal{N}_0$ distributes a portion of a quantum state among external nodes $\mathcal{N}_1, \ldots, \mathcal{N}_k$ (in this depiction, $k=4$). This distribution is done through quantum channels and thus may be vulnerable to a malicious eavesdropper, whose (potential) interaction is depicted with a red ring. To ensure a sense of security, the trusted nodes can adopt the trap code since the decryption operations are all performed locally.}
\label{fig:QASMultipartite}
\end{figure}

One could also consider generalizing the protocol to a multipartite scenario, illustrated in Figure~\ref{fig:QASMultipartite}. This would be a practical tool for any spatially distributed quantum metrology scheme \cite{komar2014quantum, ge2018, proctor2018, rubio2020}.
Here, a central node is connected to external nodes via quantum channels, which may be simultaneously intercepted by a malicious adversary. The central node sends a portion of an entangled quantum state to each of the external nodes, after which the external nodes encode a local parameter on their portion of the quantum state for a spatially distributed quantum metrology scheme. The trap code can be adopted in this spatially distributed and multipartite setting since the decryption operations are local, and thus recover the same notions of privacy and soundness.

In this article we devised an estimation strategy for a general quantum metrology problem in the presence of the a malicious adversary. We quantify how the uncertainty in the resource biases the estimate of the unknown parameter and decreases the overall precision. These results are similar in flavour to those of noisy quantum metrology \cite{escher2011a, demkowicz2012}; however, crucially, here we make no assumption on the type of `noise' or that it acts honestly. Of course, for the estimation results to be practical, we must certify that the overall resource is at least similar to the resource in the ideal setting sans malicious adversary, given by Eq.~\eqref{eqn:bound}. We demonstrate that such a statement can be made by incorporating a cryptographic protocol with a necessary notion of soundness.

The cryptographic scenario that we consider is when the underlying quantum metrology problem utilizes an unsecured quantum channel, similar to that of the work in \cite{huang2019}. We present two completely private protocols, which differ in practicability and efficiency. Although the Clifford code is more efficient, the required entanglement is highly impractical. In contrast, the trap code is only slightly more demanding than nonsecure versions, requiring only local Clifford operations for encryption. Although the results are derived with single parameter quantum metrology in mind, the methodology used can be easily extended to multiparameter estimation strategies \cite{szczykulska2016,ragy2016}. Similarly, we assumed a specific (common) estimation strategy; nonetheless, the mathematical methodology can be adapted to other estimation strategies.

\textbf{Acknowledgments.} We acknowledge the support of the European Unions Horizon 2020 Research and Innovation Programme under Grant Agreement No. 820445 (QIA), and the ANR through the ANR-17-CE24-0035 VanQuTe.

\bibliography{SecureChannelMetrology}

\begin{thebibliography}{30}%
\makeatletter
\providecommand \@ifxundefined [1]{%
 \@ifx{#1\undefined}
}%
\providecommand \@ifnum [1]{%
 \ifnum #1\expandafter \@firstoftwo
 \else \expandafter \@secondoftwo
 \fi
}%
\providecommand \@ifx [1]{%
 \ifx #1\expandafter \@firstoftwo
 \else \expandafter \@secondoftwo
 \fi
}%
\providecommand \natexlab [1]{#1}%
\providecommand \enquote  [1]{``#1''}%
\providecommand \bibnamefont  [1]{#1}%
\providecommand \bibfnamefont [1]{#1}%
\providecommand \citenamefont [1]{#1}%
\providecommand \href@noop [0]{\@secondoftwo}%
\providecommand \href [0]{\begingroup \@sanitize@url \@href}%
\providecommand \@href[1]{\@@startlink{#1}\@@href}%
\providecommand \@@href[1]{\endgroup#1\@@endlink}%
\providecommand \@sanitize@url [0]{\catcode `\\12\catcode `\$12\catcode
  `\&12\catcode `\#12\catcode `\^12\catcode `\_12\catcode `\%12\relax}%
\providecommand \@@startlink[1]{}%
\providecommand \@@endlink[0]{}%
\providecommand \url  [0]{\begingroup\@sanitize@url \@url }%
\providecommand \@url [1]{\endgroup\@href {#1}{\urlprefix }}%
\providecommand \urlprefix  [0]{URL }%
\providecommand \Eprint [0]{\href }%
\providecommand \doibase [0]{https://doi.org/}%
\providecommand \selectlanguage [0]{\@gobble}%
\providecommand \bibinfo  [0]{\@secondoftwo}%
\providecommand \bibfield  [0]{\@secondoftwo}%
\providecommand \translation [1]{[#1]}%
\providecommand \BibitemOpen [0]{}%
\providecommand \bibitemStop [0]{}%
\providecommand \bibitemNoStop [0]{.\EOS\space}%
\providecommand \EOS [0]{\spacefactor3000\relax}%
\providecommand \BibitemShut  [1]{\csname bibitem#1\endcsname}%
\let\auto@bib@innerbib\@empty
%</preamble>
\bibitem [{\citenamefont {Giovannetti}\ \emph {et~al.}(2006)\citenamefont
  {Giovannetti}, \citenamefont {Lloyd},\ and\ \citenamefont
  {Maccone}}]{giovannetti2006}%
  \BibitemOpen
  \bibfield  {author} {\bibinfo {author} {\bibfnamefont {V.}~\bibnamefont
  {Giovannetti}}, \bibinfo {author} {\bibfnamefont {S.}~\bibnamefont {Lloyd}},\
  and\ \bibinfo {author} {\bibfnamefont {L.}~\bibnamefont {Maccone}},\
  }\bibfield  {title} {\bibinfo {title} {Quantum metrology},\ }\href@noop {}
  {\bibfield  {journal} {\bibinfo  {journal} {Physical Review Letters}\
  }\textbf {\bibinfo {volume} {96}},\ \bibinfo {pages} {010401} (\bibinfo
  {year} {2006})}\BibitemShut {NoStop}%
\bibitem [{\citenamefont {Giovannetti}\ \emph {et~al.}(2011)\citenamefont
  {Giovannetti}, \citenamefont {Lloyd},\ and\ \citenamefont
  {Maccone}}]{giovannetti2011}%
  \BibitemOpen
  \bibfield  {author} {\bibinfo {author} {\bibfnamefont {V.}~\bibnamefont
  {Giovannetti}}, \bibinfo {author} {\bibfnamefont {S.}~\bibnamefont {Lloyd}},\
  and\ \bibinfo {author} {\bibfnamefont {L.}~\bibnamefont {Maccone}},\
  }\bibfield  {title} {\bibinfo {title} {Advances in quantum metrology},\
  }\href@noop {} {\bibfield  {journal} {\bibinfo  {journal} {Nature Photonics}\
  }\textbf {\bibinfo {volume} {5}},\ \bibinfo {pages} {222} (\bibinfo {year}
  {2011})}\BibitemShut {NoStop}%
\bibitem [{\citenamefont {Pirandola}\ \emph {et~al.}(2020)\citenamefont
  {Pirandola}, \citenamefont {Andersen}, \citenamefont {Banchi}, \citenamefont
  {Berta}, \citenamefont {Bunandar}, \citenamefont {Colbeck}, \citenamefont
  {Englund}, \citenamefont {Gehring}, \citenamefont {Lupo}, \citenamefont
  {Ottaviani} \emph {et~al.}}]{pirandola2020advances}%
  \BibitemOpen
  \bibfield  {author} {\bibinfo {author} {\bibfnamefont {S.}~\bibnamefont
  {Pirandola}}, \bibinfo {author} {\bibfnamefont {U.~L.}\ \bibnamefont
  {Andersen}}, \bibinfo {author} {\bibfnamefont {L.}~\bibnamefont {Banchi}},
  \bibinfo {author} {\bibfnamefont {M.}~\bibnamefont {Berta}}, \bibinfo
  {author} {\bibfnamefont {D.}~\bibnamefont {Bunandar}}, \bibinfo {author}
  {\bibfnamefont {R.}~\bibnamefont {Colbeck}}, \bibinfo {author} {\bibfnamefont
  {D.}~\bibnamefont {Englund}}, \bibinfo {author} {\bibfnamefont
  {T.}~\bibnamefont {Gehring}}, \bibinfo {author} {\bibfnamefont
  {C.}~\bibnamefont {Lupo}}, \bibinfo {author} {\bibfnamefont {C.}~\bibnamefont
  {Ottaviani}}, \emph {et~al.},\ }\bibfield  {title} {\bibinfo {title}
  {Advances in quantum cryptography},\ }\href@noop {} {\bibfield  {journal}
  {\bibinfo  {journal} {Advances in Optics and Photonics}\ }\textbf {\bibinfo
  {volume} {12}},\ \bibinfo {pages} {1012} (\bibinfo {year}
  {2020})}\BibitemShut {NoStop}%
\bibitem [{\citenamefont {Wehner}\ \emph {et~al.}(2018)\citenamefont {Wehner},
  \citenamefont {Elkouss},\ and\ \citenamefont {Hanson}}]{wehner2018quantum}%
  \BibitemOpen
  \bibfield  {author} {\bibinfo {author} {\bibfnamefont {S.}~\bibnamefont
  {Wehner}}, \bibinfo {author} {\bibfnamefont {D.}~\bibnamefont {Elkouss}},\
  and\ \bibinfo {author} {\bibfnamefont {R.}~\bibnamefont {Hanson}},\
  }\bibfield  {title} {\bibinfo {title} {Quantum internet: A vision for the
  road ahead},\ }\href@noop {} {\bibfield  {journal} {\bibinfo  {journal}
  {Science}\ }\textbf {\bibinfo {volume} {362}},\ \bibinfo {pages} {eaam9288}
  (\bibinfo {year} {2018})}\BibitemShut {NoStop}%
\bibitem [{\citenamefont {Komar}\ \emph {et~al.}(2014)\citenamefont {Komar},
  \citenamefont {Kessler}, \citenamefont {Bishof}, \citenamefont {Jiang},
  \citenamefont {S{\o}rensen}, \citenamefont {Ye},\ and\ \citenamefont
  {Lukin}}]{komar2014quantum}%
  \BibitemOpen
  \bibfield  {author} {\bibinfo {author} {\bibfnamefont {P.}~\bibnamefont
  {Komar}}, \bibinfo {author} {\bibfnamefont {E.~M.}\ \bibnamefont {Kessler}},
  \bibinfo {author} {\bibfnamefont {M.}~\bibnamefont {Bishof}}, \bibinfo
  {author} {\bibfnamefont {L.}~\bibnamefont {Jiang}}, \bibinfo {author}
  {\bibfnamefont {A.~S.}\ \bibnamefont {S{\o}rensen}}, \bibinfo {author}
  {\bibfnamefont {J.}~\bibnamefont {Ye}},\ and\ \bibinfo {author}
  {\bibfnamefont {M.~D.}\ \bibnamefont {Lukin}},\ }\bibfield  {title} {\bibinfo
  {title} {A quantum network of clocks},\ }\href@noop {} {\bibfield  {journal}
  {\bibinfo  {journal} {Nature Physics}\ }\textbf {\bibinfo {volume} {10}},\
  \bibinfo {pages} {582} (\bibinfo {year} {2014})}\BibitemShut {NoStop}%
\bibitem [{\citenamefont {Huang}\ \emph {et~al.}(2019)\citenamefont {Huang},
  \citenamefont {Macchiavello},\ and\ \citenamefont {Maccone}}]{huang2019}%
  \BibitemOpen
  \bibfield  {author} {\bibinfo {author} {\bibfnamefont {Z.}~\bibnamefont
  {Huang}}, \bibinfo {author} {\bibfnamefont {C.}~\bibnamefont
  {Macchiavello}},\ and\ \bibinfo {author} {\bibfnamefont {L.}~\bibnamefont
  {Maccone}},\ }\bibfield  {title} {\bibinfo {title} {Cryptographic quantum
  metrology},\ }\href@noop {} {\bibfield  {journal} {\bibinfo  {journal}
  {Physical Review A}\ }\textbf {\bibinfo {volume} {99}},\ \bibinfo {pages}
  {022314} (\bibinfo {year} {2019})}\BibitemShut {NoStop}%
\bibitem [{\citenamefont {Xie}\ \emph {et~al.}(2018)\citenamefont {Xie},
  \citenamefont {Xu}, \citenamefont {Chen},\ and\ \citenamefont
  {Wang}}]{xie2018}%
  \BibitemOpen
  \bibfield  {author} {\bibinfo {author} {\bibfnamefont {D.}~\bibnamefont
  {Xie}}, \bibinfo {author} {\bibfnamefont {C.}~\bibnamefont {Xu}}, \bibinfo
  {author} {\bibfnamefont {J.}~\bibnamefont {Chen}},\ and\ \bibinfo {author}
  {\bibfnamefont {A.~M.}\ \bibnamefont {Wang}},\ }\bibfield  {title} {\bibinfo
  {title} {High-dimensional cryptographic quantum parameter estimation},\
  }\href@noop {} {\bibfield  {journal} {\bibinfo  {journal} {Quantum
  Information Processing}\ }\textbf {\bibinfo {volume} {17}},\ \bibinfo {pages}
  {116} (\bibinfo {year} {2018})}\BibitemShut {NoStop}%
\bibitem [{\citenamefont {Takeuchi}\ \emph {et~al.}(2019)\citenamefont
  {Takeuchi}, \citenamefont {Matsuzaki}, \citenamefont {Miyanishi},
  \citenamefont {Sugiyama},\ and\ \citenamefont {Munro}}]{takeuchi2019}%
  \BibitemOpen
  \bibfield  {author} {\bibinfo {author} {\bibfnamefont {Y.}~\bibnamefont
  {Takeuchi}}, \bibinfo {author} {\bibfnamefont {Y.}~\bibnamefont {Matsuzaki}},
  \bibinfo {author} {\bibfnamefont {K.}~\bibnamefont {Miyanishi}}, \bibinfo
  {author} {\bibfnamefont {T.}~\bibnamefont {Sugiyama}},\ and\ \bibinfo
  {author} {\bibfnamefont {W.~J.}\ \bibnamefont {Munro}},\ }\bibfield  {title}
  {\bibinfo {title} {Quantum remote sensing with asymmetric information gain},\
  }\href@noop {} {\bibfield  {journal} {\bibinfo  {journal} {Physical Review
  A}\ }\textbf {\bibinfo {volume} {99}},\ \bibinfo {pages} {022325} (\bibinfo
  {year} {2019})}\BibitemShut {NoStop}%
\bibitem [{\citenamefont {Okane}\ \emph {et~al.}(2021)\citenamefont {Okane},
  \citenamefont {Hakoshima}, \citenamefont {Takeuchi}, \citenamefont {Seki},\
  and\ \citenamefont {Matsuzaki}}]{okane2020}%
  \BibitemOpen
  \bibfield  {author} {\bibinfo {author} {\bibfnamefont {H.}~\bibnamefont
  {Okane}}, \bibinfo {author} {\bibfnamefont {H.}~\bibnamefont {Hakoshima}},
  \bibinfo {author} {\bibfnamefont {Y.}~\bibnamefont {Takeuchi}}, \bibinfo
  {author} {\bibfnamefont {Y.}~\bibnamefont {Seki}},\ and\ \bibinfo {author}
  {\bibfnamefont {Y.}~\bibnamefont {Matsuzaki}},\ }\bibfield  {title} {\bibinfo
  {title} {Quantum remote sensing under the effect of dephasing},\ }\href@noop
  {} {\bibfield  {journal} {\bibinfo  {journal} {Physical Review A}\ }\textbf
  {\bibinfo {volume} {104}},\ \bibinfo {pages} {062610} (\bibinfo {year}
  {2021})}\BibitemShut {NoStop}%
\bibitem [{\citenamefont {Yin}\ \emph {et~al.}(2020)\citenamefont {Yin},
  \citenamefont {Takeuchi}, \citenamefont {Zhang}, \citenamefont {Yin},
  \citenamefont {Matsuzaki}, \citenamefont {Peng}, \citenamefont {Xu},
  \citenamefont {Xu}, \citenamefont {Tang}, \citenamefont {Zhou} \emph
  {et~al.}}]{yin2020}%
  \BibitemOpen
  \bibfield  {author} {\bibinfo {author} {\bibfnamefont {P.}~\bibnamefont
  {Yin}}, \bibinfo {author} {\bibfnamefont {Y.}~\bibnamefont {Takeuchi}},
  \bibinfo {author} {\bibfnamefont {W.-H.}\ \bibnamefont {Zhang}}, \bibinfo
  {author} {\bibfnamefont {Z.-Q.}\ \bibnamefont {Yin}}, \bibinfo {author}
  {\bibfnamefont {Y.}~\bibnamefont {Matsuzaki}}, \bibinfo {author}
  {\bibfnamefont {X.-X.}\ \bibnamefont {Peng}}, \bibinfo {author}
  {\bibfnamefont {X.-Y.}\ \bibnamefont {Xu}}, \bibinfo {author} {\bibfnamefont
  {J.-S.}\ \bibnamefont {Xu}}, \bibinfo {author} {\bibfnamefont {J.-S.}\
  \bibnamefont {Tang}}, \bibinfo {author} {\bibfnamefont {Z.-Q.}\ \bibnamefont
  {Zhou}}, \emph {et~al.},\ }\bibfield  {title} {\bibinfo {title} {Experimental
  demonstration of secure quantum remote sensing},\ }\href@noop {} {\bibfield
  {journal} {\bibinfo  {journal} {Physical Review Applied}\ }\textbf {\bibinfo
  {volume} {14}},\ \bibinfo {pages} {014065} (\bibinfo {year}
  {2020})}\BibitemShut {NoStop}%
\bibitem [{\citenamefont {Braunstein}\ and\ \citenamefont
  {Caves}(1994)}]{braunstein1994}%
  \BibitemOpen
  \bibfield  {author} {\bibinfo {author} {\bibfnamefont {S.~L.}\ \bibnamefont
  {Braunstein}}\ and\ \bibinfo {author} {\bibfnamefont {C.~M.}\ \bibnamefont
  {Caves}},\ }\bibfield  {title} {\bibinfo {title} {Statistical distance and
  the geometry of quantum states},\ }\href@noop {} {\bibfield  {journal}
  {\bibinfo  {journal} {Physical Review Letters}\ }\textbf {\bibinfo {volume}
  {72}},\ \bibinfo {pages} {3439} (\bibinfo {year} {1994})}\BibitemShut
  {NoStop}%
\bibitem [{\citenamefont {Luo}(2000)}]{luo2000}%
  \BibitemOpen
  \bibfield  {author} {\bibinfo {author} {\bibfnamefont {S.}~\bibnamefont
  {Luo}},\ }\bibfield  {title} {\bibinfo {title} {Quantum fisher information
  and uncertainty relations},\ }\href@noop {} {\bibfield  {journal} {\bibinfo
  {journal} {Letters in Mathematical Physics}\ }\textbf {\bibinfo {volume}
  {53}},\ \bibinfo {pages} {243} (\bibinfo {year} {2000})}\BibitemShut
  {NoStop}%
\bibitem [{\citenamefont {Escher}\ \emph {et~al.}(2011)\citenamefont {Escher},
  \citenamefont {de~Matos~Filho},\ and\ \citenamefont
  {Davidovich}}]{escher2011a}%
  \BibitemOpen
  \bibfield  {author} {\bibinfo {author} {\bibfnamefont {B.}~\bibnamefont
  {Escher}}, \bibinfo {author} {\bibfnamefont {R.}~\bibnamefont
  {de~Matos~Filho}},\ and\ \bibinfo {author} {\bibfnamefont {L.}~\bibnamefont
  {Davidovich}},\ }\bibfield  {title} {\bibinfo {title} {Quantum metrology for
  noisy systems},\ }\href@noop {} {\bibfield  {journal} {\bibinfo  {journal}
  {Brazilian Journal of Physics}\ }\textbf {\bibinfo {volume} {41}},\ \bibinfo
  {pages} {229} (\bibinfo {year} {2011})}\BibitemShut {NoStop}%
\bibitem [{\citenamefont {Demkowicz-Dobrza{\'{n}}ski}\ \emph
  {et~al.}(2012)\citenamefont {Demkowicz-Dobrza{\'{n}}ski}, \citenamefont
  {Ko{\l}ody{\'{n}}ski},\ and\ \citenamefont
  {Gu{\c{t}}{\u{a}}}}]{demkowicz2012}%
  \BibitemOpen
  \bibfield  {author} {\bibinfo {author} {\bibfnamefont {R.}~\bibnamefont
  {Demkowicz-Dobrza{\'{n}}ski}}, \bibinfo {author} {\bibfnamefont
  {J.}~\bibnamefont {Ko{\l}ody{\'{n}}ski}},\ and\ \bibinfo {author}
  {\bibfnamefont {M.}~\bibnamefont {Gu{\c{t}}{\u{a}}}},\ }\bibfield  {title}
  {\bibinfo {title} {The elusive heisenberg limit in quantum-enhanced
  metrology},\ }\href@noop {} {\bibfield  {journal} {\bibinfo  {journal}
  {Nature Communications}\ }\textbf {\bibinfo {volume} {3}},\ \bibinfo {pages}
  {1} (\bibinfo {year} {2012})}\BibitemShut {NoStop}%
\bibitem [{\citenamefont {Ge}\ \emph {et~al.}(2018)\citenamefont {Ge},
  \citenamefont {Jacobs}, \citenamefont {Eldredge}, \citenamefont {Gorshkov},\
  and\ \citenamefont {Foss-Feig}}]{ge2018}%
  \BibitemOpen
  \bibfield  {author} {\bibinfo {author} {\bibfnamefont {W.}~\bibnamefont
  {Ge}}, \bibinfo {author} {\bibfnamefont {K.}~\bibnamefont {Jacobs}}, \bibinfo
  {author} {\bibfnamefont {Z.}~\bibnamefont {Eldredge}}, \bibinfo {author}
  {\bibfnamefont {A.~V.}\ \bibnamefont {Gorshkov}},\ and\ \bibinfo {author}
  {\bibfnamefont {M.}~\bibnamefont {Foss-Feig}},\ }\bibfield  {title} {\bibinfo
  {title} {Distributed quantum metrology with linear networks and separable
  inputs},\ }\href@noop {} {\bibfield  {journal} {\bibinfo  {journal} {Physical
  Review Letters}\ }\textbf {\bibinfo {volume} {121}},\ \bibinfo {pages}
  {043604} (\bibinfo {year} {2018})}\BibitemShut {NoStop}%
\bibitem [{\citenamefont {Proctor}\ \emph {et~al.}(2018)\citenamefont
  {Proctor}, \citenamefont {Knott},\ and\ \citenamefont
  {Dunningham}}]{proctor2018}%
  \BibitemOpen
  \bibfield  {author} {\bibinfo {author} {\bibfnamefont {T.~J.}\ \bibnamefont
  {Proctor}}, \bibinfo {author} {\bibfnamefont {P.~A.}\ \bibnamefont {Knott}},\
  and\ \bibinfo {author} {\bibfnamefont {J.~A.}\ \bibnamefont {Dunningham}},\
  }\bibfield  {title} {\bibinfo {title} {Multiparameter estimation in networked
  quantum sensors},\ }\href@noop {} {\bibfield  {journal} {\bibinfo  {journal}
  {Physical review letters}\ }\textbf {\bibinfo {volume} {120}},\ \bibinfo
  {pages} {080501} (\bibinfo {year} {2018})}\BibitemShut {NoStop}%
\bibitem [{\citenamefont {Rubio}\ \emph {et~al.}(2020)\citenamefont {Rubio},
  \citenamefont {Knott}, \citenamefont {Proctor},\ and\ \citenamefont
  {Dunningham}}]{rubio2020}%
  \BibitemOpen
  \bibfield  {author} {\bibinfo {author} {\bibfnamefont {J.}~\bibnamefont
  {Rubio}}, \bibinfo {author} {\bibfnamefont {P.~A.}\ \bibnamefont {Knott}},
  \bibinfo {author} {\bibfnamefont {T.~J.}\ \bibnamefont {Proctor}},\ and\
  \bibinfo {author} {\bibfnamefont {J.~A.}\ \bibnamefont {Dunningham}},\
  }\bibfield  {title} {\bibinfo {title} {Quantum sensing networks for the
  estimation of linear functions},\ }\href@noop {} {\bibfield  {journal}
  {\bibinfo  {journal} {Journal of Physics A: Mathematical and Theoretical}\
  }\textbf {\bibinfo {volume} {53}},\ \bibinfo {pages} {344001} (\bibinfo
  {year} {2020})}\BibitemShut {NoStop}%
\bibitem [{\citenamefont {T{\'o}th}\ and\ \citenamefont
  {Apellaniz}(2014)}]{toth2014}%
  \BibitemOpen
  \bibfield  {author} {\bibinfo {author} {\bibfnamefont {G.}~\bibnamefont
  {T{\'o}th}}\ and\ \bibinfo {author} {\bibfnamefont {I.}~\bibnamefont
  {Apellaniz}},\ }\bibfield  {title} {\bibinfo {title} {Quantum metrology from
  a quantum information science perspective},\ }\href@noop {} {\bibfield
  {journal} {\bibinfo  {journal} {Journal of Physics A: Mathematical and
  Theoretical}\ }\textbf {\bibinfo {volume} {47}},\ \bibinfo {pages} {424006}
  (\bibinfo {year} {2014})}\BibitemShut {NoStop}%
\bibitem [{\citenamefont {Nielsen}\ and\ \citenamefont {Chuang}(2000)}]{NC}%
  \BibitemOpen
  \bibfield  {author} {\bibinfo {author} {\bibfnamefont {M.~A.}\ \bibnamefont
  {Nielsen}}\ and\ \bibinfo {author} {\bibfnamefont {I.~L.}\ \bibnamefont
  {Chuang}},\ }\href@noop {} {\emph {\bibinfo {title} {Quantum Computation and
  Quantum Information}}}\ (\bibinfo  {publisher} {Cambridge University Press},\
  \bibinfo {year} {2000})\BibitemShut {NoStop}%
\bibitem [{\citenamefont {Barnum}\ \emph {et~al.}(2002)\citenamefont {Barnum},
  \citenamefont {Cr{\'e}peau}, \citenamefont {Gottesman}, \citenamefont
  {Smith},\ and\ \citenamefont {Tapp}}]{barnum2002}%
  \BibitemOpen
  \bibfield  {author} {\bibinfo {author} {\bibfnamefont {H.}~\bibnamefont
  {Barnum}}, \bibinfo {author} {\bibfnamefont {C.}~\bibnamefont {Cr{\'e}peau}},
  \bibinfo {author} {\bibfnamefont {D.}~\bibnamefont {Gottesman}}, \bibinfo
  {author} {\bibfnamefont {A.}~\bibnamefont {Smith}},\ and\ \bibinfo {author}
  {\bibfnamefont {A.}~\bibnamefont {Tapp}},\ }\bibfield  {title} {\bibinfo
  {title} {Authentication of quantum messages},\ }in\ \href@noop {} {\emph
  {\bibinfo {booktitle} {The 43rd Annual IEEE Symposium on Foundations of
  Computer Science, 2002. Proceedings.}}}\ (\bibinfo {organization} {IEEE},\
  \bibinfo {year} {2002})\ pp.\ \bibinfo {pages} {449--458}\BibitemShut
  {NoStop}%
\bibitem [{\citenamefont {Broadbent}\ and\ \citenamefont
  {Wainewright}(2016)}]{broadbent2016}%
  \BibitemOpen
  \bibfield  {author} {\bibinfo {author} {\bibfnamefont {A.}~\bibnamefont
  {Broadbent}}\ and\ \bibinfo {author} {\bibfnamefont {E.}~\bibnamefont
  {Wainewright}},\ }\bibfield  {title} {\bibinfo {title} {Efficient simulation
  for quantum message authentication},\ }in\ \href@noop {} {\emph {\bibinfo
  {booktitle} {International Conference on Information Theoretic Security}}},\
  \bibinfo {editor} {edited by\ \bibinfo {editor} {\bibfnamefont {A.~C.}\
  \bibnamefont {Nascimento}}\ and\ \bibinfo {editor} {\bibfnamefont
  {P.}~\bibnamefont {Barreto}}}\ (\bibinfo {organization} {Springer},\ \bibinfo
  {address} {New York},\ \bibinfo {year} {2016})\ pp.\ \bibinfo {pages}
  {72--91}\BibitemShut {NoStop}%
\bibitem [{\citenamefont {Gheorghiu}\ \emph {et~al.}(2019)\citenamefont
  {Gheorghiu}, \citenamefont {Kapourniotis},\ and\ \citenamefont
  {Kashefi}}]{gheorghiu2019}%
  \BibitemOpen
  \bibfield  {author} {\bibinfo {author} {\bibfnamefont {A.}~\bibnamefont
  {Gheorghiu}}, \bibinfo {author} {\bibfnamefont {T.}~\bibnamefont
  {Kapourniotis}},\ and\ \bibinfo {author} {\bibfnamefont {E.}~\bibnamefont
  {Kashefi}},\ }\bibfield  {title} {\bibinfo {title} {Verification of quantum
  computation: An overview of existing approaches},\ }\href@noop {} {\bibfield
  {journal} {\bibinfo  {journal} {Theory of computing systems}\ }\textbf
  {\bibinfo {volume} {63}},\ \bibinfo {pages} {715} (\bibinfo {year}
  {2019})}\BibitemShut {NoStop}%
\bibitem [{\citenamefont {Broadbent}\ \emph {et~al.}(2013)\citenamefont
  {Broadbent}, \citenamefont {Gutoski},\ and\ \citenamefont
  {Stebila}}]{broadbent2013}%
  \BibitemOpen
  \bibfield  {author} {\bibinfo {author} {\bibfnamefont {A.}~\bibnamefont
  {Broadbent}}, \bibinfo {author} {\bibfnamefont {G.}~\bibnamefont {Gutoski}},\
  and\ \bibinfo {author} {\bibfnamefont {D.}~\bibnamefont {Stebila}},\
  }\bibfield  {title} {\bibinfo {title} {Quantum one-time programs},\ }in\
  \href@noop {} {\emph {\bibinfo {booktitle} {Annual Cryptology Conference}}},\
  \bibinfo {editor} {edited by\ \bibinfo {editor} {\bibfnamefont
  {R.}~\bibnamefont {Canetti}}\ and\ \bibinfo {editor} {\bibfnamefont {J.~A.}\
  \bibnamefont {Garay}}}\ (\bibinfo {organization} {Springer},\ \bibinfo
  {address} {Berlin},\ \bibinfo {year} {2013})\ pp.\ \bibinfo {pages}
  {344--360}\BibitemShut {NoStop}%
\bibitem [{\citenamefont {Aharonov}\ \emph {et~al.}(2017)\citenamefont
  {Aharonov}, \citenamefont {Ben-Or}, \citenamefont {Eban},\ and\ \citenamefont
  {Mahadev}}]{aharonov2017}%
  \BibitemOpen
  \bibfield  {author} {\bibinfo {author} {\bibfnamefont {D.}~\bibnamefont
  {Aharonov}}, \bibinfo {author} {\bibfnamefont {M.}~\bibnamefont {Ben-Or}},
  \bibinfo {author} {\bibfnamefont {E.}~\bibnamefont {Eban}},\ and\ \bibinfo
  {author} {\bibfnamefont {U.}~\bibnamefont {Mahadev}},\ }\bibfield  {title}
  {\bibinfo {title} {Interactive proofs for quantum computations},\ }\href@noop
  {} {\bibfield  {journal} {\bibinfo  {journal} {arXiv preprint
  arXiv:1704.04487}\ } (\bibinfo {year} {2017})}\BibitemShut {NoStop}%
\bibitem [{\citenamefont {Dankert}\ \emph {et~al.}(2009)\citenamefont
  {Dankert}, \citenamefont {Cleve}, \citenamefont {Emerson},\ and\
  \citenamefont {Livine}}]{dankert2009}%
  \BibitemOpen
  \bibfield  {author} {\bibinfo {author} {\bibfnamefont {C.}~\bibnamefont
  {Dankert}}, \bibinfo {author} {\bibfnamefont {R.}~\bibnamefont {Cleve}},
  \bibinfo {author} {\bibfnamefont {J.}~\bibnamefont {Emerson}},\ and\ \bibinfo
  {author} {\bibfnamefont {E.}~\bibnamefont {Livine}},\ }\bibfield  {title}
  {\bibinfo {title} {Exact and approximate unitary 2-designs and their
  application to fidelity estimation},\ }\href@noop {} {\bibfield  {journal}
  {\bibinfo  {journal} {Physical Review A}\ }\textbf {\bibinfo {volume} {80}},\
  \bibinfo {pages} {012304} (\bibinfo {year} {2009})}\BibitemShut {NoStop}%
\bibitem [{not()}]{note1}%
  \BibitemOpen
  \href@noop {} {\bibinfo {title} {If we consider the attack where eve replaces
  the quantum state with the maximally mixed state, then $p_\text{acc}$ is
  small but non-zero, and the resulting state $\rho^\prime$ is useless for
  quantum metrology.}}\BibitemShut {Stop}%
\bibitem [{\citenamefont {Zhu}\ and\ \citenamefont {Hayashi}(2019)}]{zhu2019}%
  \BibitemOpen
  \bibfield  {author} {\bibinfo {author} {\bibfnamefont {H.}~\bibnamefont
  {Zhu}}\ and\ \bibinfo {author} {\bibfnamefont {M.}~\bibnamefont {Hayashi}},\
  }\bibfield  {title} {\bibinfo {title} {General framework for verifying pure
  quantum states in the adversarial scenario},\ }\href@noop {} {\bibfield
  {journal} {\bibinfo  {journal} {Physical Review A}\ }\textbf {\bibinfo
  {volume} {100}},\ \bibinfo {pages} {062335} (\bibinfo {year}
  {2019})}\BibitemShut {NoStop}%
\bibitem [{\citenamefont {Fuchs}\ and\ \citenamefont {Van
  De~Graaf}(1999)}]{fuchs1999}%
  \BibitemOpen
  \bibfield  {author} {\bibinfo {author} {\bibfnamefont {C.~A.}\ \bibnamefont
  {Fuchs}}\ and\ \bibinfo {author} {\bibfnamefont {J.}~\bibnamefont {Van
  De~Graaf}},\ }\bibfield  {title} {\bibinfo {title} {Cryptographic
  distinguishability measures for quantum-mechanical states},\ }\href@noop {}
  {\bibfield  {journal} {\bibinfo  {journal} {IEEE Transactions on Information
  Theory}\ }\textbf {\bibinfo {volume} {45}},\ \bibinfo {pages} {1216}
  (\bibinfo {year} {1999})}\BibitemShut {NoStop}%
\bibitem [{\citenamefont {Szczykulska}\ \emph {et~al.}(2016)\citenamefont
  {Szczykulska}, \citenamefont {Baumgratz},\ and\ \citenamefont
  {Datta}}]{szczykulska2016}%
  \BibitemOpen
  \bibfield  {author} {\bibinfo {author} {\bibfnamefont {M.}~\bibnamefont
  {Szczykulska}}, \bibinfo {author} {\bibfnamefont {T.}~\bibnamefont
  {Baumgratz}},\ and\ \bibinfo {author} {\bibfnamefont {A.}~\bibnamefont
  {Datta}},\ }\bibfield  {title} {\bibinfo {title} {Multi-parameter quantum
  metrology},\ }\href@noop {} {\bibfield  {journal} {\bibinfo  {journal}
  {Advances in Physics: X}\ }\textbf {\bibinfo {volume} {1}},\ \bibinfo {pages}
  {621} (\bibinfo {year} {2016})}\BibitemShut {NoStop}%
\bibitem [{\citenamefont {Ragy}\ \emph {et~al.}(2016)\citenamefont {Ragy},
  \citenamefont {Jarzyna},\ and\ \citenamefont
  {Demkowicz-Dobrza{\'n}ski}}]{ragy2016}%
  \BibitemOpen
  \bibfield  {author} {\bibinfo {author} {\bibfnamefont {S.}~\bibnamefont
  {Ragy}}, \bibinfo {author} {\bibfnamefont {M.}~\bibnamefont {Jarzyna}},\ and\
  \bibinfo {author} {\bibfnamefont {R.}~\bibnamefont
  {Demkowicz-Dobrza{\'n}ski}},\ }\bibfield  {title} {\bibinfo {title}
  {Compatibility in multiparameter quantum metrology},\ }\href@noop {}
  {\bibfield  {journal} {\bibinfo  {journal} {Physical Review A}\ }\textbf
  {\bibinfo {volume} {94}},\ \bibinfo {pages} {052108} (\bibinfo {year}
  {2016})}\BibitemShut {NoStop}%
\end{thebibliography}%

\appendix

\onecolumngrid

\section{Appendix A: Privacy of the Protocols}

\setcounter{equation}{0}
\renewcommand\theequation{A.\arabic{equation}}

Recall that a protocol is completely private if an eavesdropper interacting with the quantum channel cannot distinguish the quantum state from the maximally mixed state. Without loss of generality, we can write that the quantum state being encrypted by Alice or Bob is some $m$ qubit quantum state
\begin{equation}
\rho=2^{-m} \sum_{P \in \mathcal{P}_m} \Tr(\rho P) P,
\end{equation}
where $\mathcal{P}_m=\{ \mathbb{I},X,Y,Z \}^{\otimes m}$ is the $m$ dimensional Pauli group. When using the Clifford code, the effective state viewed by an eavesdropper is
\begin{equation}
\rho_\text{eff}=2^{-m} |\mathcal{C}_m|^{-1} \sum_{C \in \mathcal{C}_m} \sum_{P \in \mathcal{P}_m} \Tr(\rho P) CPC^{\dagger}.
\end{equation}
For every $P \neq \mathbb{I}$, the sum over the Clifford group can be broken into pairs $(C,C^\prime)$ of operators such that $CPC^\dagger=-C^\prime P C^{\prime \dagger}$. Thus the only non vanishing term is $P = \mathbb{I}$ and
\begin{equation}
\rho_\text{eff}=2^{-m} \Tr(\rho \mathbb{I}) \mathbb{I} = \mathbb{I} /2^{m}.
\end{equation}

Using similar logic, the same result can be shown when using the trap code encryption. By decomposing $P=P_1 \otimes \ldots \otimes P_m$, where each $P_i \in \mathcal{P}_1$, one can write that the effective state viewed by an eavesdropper is
\begin{equation}
\rho_\text{eff}=2^{-m} |\mathcal{C}_1^{\otimes m}|^{-1} \sum_{P \in \mathcal{P}_m} \Tr(\rho P) \bigotimes_{i=1}^m \sum_{C \in \mathcal{C}_1} CP_iC^{\dagger}.
\end{equation}
By the same intuition, the only non-vanishing term is when $P_1=\ldots=P_m=\mathbb{I}$, showing that the effective state seen by an eavesdropper is the maximally mixed state.

\section{Appendix B: Soundness of the Protocols}

\setcounter{equation}{0}
\renewcommand\theequation{B.\arabic{equation}}

Prior to finding a bound for the soundness of the protocols, we first prove an analogue of the Pauli twirl lemma \cite{dankert2009}, which states that for any $m$ dimensional quantum state $\rho$, and Pauli operators $Q \neq Q^\prime \in \mathcal{P}_m$
\begin{equation}
\label{eqn:PauliTwirl}
\sum_{P \in \mathcal{P}_m} P^\dagger Q P \rho P^\dagger Q^\prime P = 0.
\end{equation}
There is a similar result known as the Clifford twirl lemma \cite{dankert2009} which states that given $Q \neq Q^\prime$
\begin{equation}
\label{eqn:CliffordTwirl}
\sum_{C \in \mathcal{C}_m} C^\dagger Q C \rho C^\dagger Q^\prime C = 0.
\end{equation}
The encryption scheme for the trap code uses operators from the group $\mathcal{C}_1^{\otimes m}$. Thus, we wish to show that
\begin{equation}
\label{eqn:1CliffordTwirl}
\sum_{C \in \mathcal{C}_1^{\otimes m}} C^\dagger Q C \rho C^\dagger Q^\prime C = 0
\end{equation}
if $Q \neq Q^\prime$. To see that the same result is obtained, one can decompose the quantum state into a sum of Pauli operators $P=P_1 \otimes \ldots \otimes P_m$ and further decomposing each single qubit Pauli $P_i$ in a diagonal form $P_i = \lambda_{P_i,1} \dyad{\lambda_{P_i,1}}+\lambda_{P_i,2} \dyad{\lambda_{P_i,2}}$. The analogous sum then can be written as
\begin{equation}
\sum_{C \in \mathcal{C}_1^{\otimes m}} C^\dagger Q_1 C \rho C^\dagger Q_2 C = 2^{-m} \sum_{P \in \mathcal{P}_m} \Tr(\rho P) \bigotimes_{i=1}^m \sum_{j=1}^2 \sum_{C \in \mathcal{C}_1} \lambda_{P_i,j}C^\dagger Q_i C \dyad{\lambda_{P_i,j}} C^\dagger Q^\prime_i C .
\end{equation}
Since $Q \neq Q^\prime$ there exists some $1 \leq i \leq m$ where $Q_i \neq Q^\prime_i$, and from the Eq.~(\ref{eqn:CliffordTwirl}) we know that
\begin{equation}
\sum_{C \in \mathcal{C}_1}C^\dagger Q_i C \dyad{\lambda_{P_i,j}} C^\dagger Q^\prime_i C=0
\end{equation}
for any $\dyad{\lambda_{P_i,j}}$. Resulting in the whole expression being equal to zero.

\subsection{The Trap Code}

We wish to find a bound the quantity
\begin{equation}
    \frac{1}{\mathcal{K}} \sum_{k \in \mathcal{K}} \Tr \big( \Pi_k \rho_\text{out} (k, \Gamma) \big),
\end{equation}
where the key $k$ encodes the choice of encoding Clifford operation $C \in \mathcal{C}_1^{\otimes n}$ and the location of the flag qubits $\vec{l}$. For all intents and purposes we can model the insertion of the flag qubits with a permutation operator $\pi$, which there are $\binom{m}{t}$ choices, acting on an initial state $\rho_0=\rho \otimes \dyad{0}^{\otimes t}$. Therefore, for a specific $k$ the output state is
\begin{equation}
    \rho_\text{out}(k, \Gamma) = C^\dagger \Gamma(C\pi \rho_0 \pi^\dagger C^\dagger) C,
\end{equation}
with corresponding projector $\Pi_k= \pi \Pi \pi^\dagger$ with
\begin{equation}
    \Pi = (\mathbb{I}-\rho) \otimes \dyad{0}^{\otimes t}.
\end{equation}
Having a fixed value for $\Pi$ greatly simplifies our computation due to the linearity of the trace
\begin{equation}
    \frac{1}{\mathcal{K}} \sum_{k \in \mathcal{K}} \Tr \big( \Pi_k \rho_\text{out} (k, \Gamma) \big) = \Tr \Big( \Pi \frac{1}{\mathcal{K}} \sum_{k \in \mathcal{K}} \pi^\dagger \rho_\text{out} \pi \Big).
\end{equation}

To begin making simplifications we use the Kraus decomposition of a CPTP map, that is $\Gamma$ can be decomposed into a sum of Kraus operators $\{ A_\alpha \}$ that satisfy the completeness relationship $\sum_\alpha A_\alpha A_\alpha^\dagger=\mathbb{I}$,
\begin{equation}
\Gamma(\rho)=\sum_\alpha A_\alpha \rho A_\alpha^\dagger.
\end{equation}
Furthermore, each Kraus operator can be expressed as a sum of Pauli operators
\begin{equation}
A_\alpha=2^{-m} \sum_{P \in \mathcal{P}_m} \Tr (A_\alpha P) P,
\end{equation}
and the completeness relation equates to 
\begin{equation}
\label{eqn:complete}
\sum_\alpha \sum_{P \in \mathcal{P}_m} |2^{-m} \Tr (A_\alpha P) |^2=1.
\end{equation}
We substitute $a_{\alpha,P} = 2^{-m} \Tr (A_\alpha P)$ for clarity. Using this substitution, we obtain
\begin{equation}
\frac{1}{|\mathcal{K}|} \sum_{k \in \mathcal{K}}  \pi^\dagger \rho_\text{out} (k,\Gamma) \pi = \frac{1}{|\mathcal{K}|} \sum_{C \in \mathcal{C}_1^{\otimes m}} \sum_{\pi} \sum_{\alpha} \sum_{P,Q \in \mathcal{P}_m} a_{\alpha,P} a_{\alpha,Q}^*  \pi^\dagger C^\dagger P C \pi \rho_0 \pi^\dagger C^\dagger Q C \pi,
\end{equation}
where an asterisk indicates the complex conjugate. At first glance this formulation is much more complicated than the original, however using the single Clifford twirl lemma, Eq.~(\ref{eqn:1CliffordTwirl}), the only non-vanishing terms occur when $P=Q$
\begin{equation}
\frac{1}{|\mathcal{K}|} \sum_{k \in \mathcal{K}}  \pi^\dagger \rho_\text{out} (k,\Gamma) \pi = \frac{1}{|\mathcal{K}|} \sum_{C \in \mathcal{C}_1^{\otimes m}} \sum_{\pi} \sum_{\alpha} \sum_{P \in \mathcal{P}_m} |a_{\alpha,P}|^2  \pi^\dagger C^\dagger P C \pi \rho_0 \pi^\dagger C^\dagger P C \pi.
\end{equation}
Next we partition $\mathcal{P}_m$ into disjoint sets $\mathcal{P}_m^{(r)}$, with $0 \leq r \leq m$, where $r$ signifies the number of non-identity terms in a Pauli, for example $\mathbb{I} \otimes X \in \mathcal{P}_2^{(1)}$, hence
\begin{equation}
\frac{1}{|\mathcal{K}|} \sum_{k \in \mathcal{K}}  \pi^\dagger \rho_\text{out} (k,\Gamma) \pi = \frac{1}{|\mathcal{K}|} \sum_{C \in \mathcal{C}_1^{\otimes m}} \sum_{\pi} \sum_{\alpha} \sum_{r=0}^m \sum_{P \in \mathcal{P}_m^{(r)}} |a_{\alpha,P}|^2  \pi^\dagger C^\dagger P C \pi \rho_0 \pi^\dagger C^\dagger P C \pi.
\end{equation}
There are $\binom{m-r}{t-s}$ choices of $\pi$ such that $s \leq r$ of the non-identity terms of $\pi^\dagger C^\dagger P  C \pi \in \mathcal{P}_m^{(r)}$ interact with $s$ of the flag qubits of $\Pi$ (and thus $r-s$ non-identity terms interact with the metrology qubits of $\Pi$). Recall that the Clifford group $C_1$ will map any $P \in \{X,Y,Z \}$ to an equal distribution over $\{ \pm X, \pm Y, \pm Z \}$. The only-non vanishing terms occur when $C^\dagger $ maps these $s$ terms exclusively onto $\pm Z$, which occurs for $3^{-s}|\mathcal{C}_1|^m$ of the local Cliffords. Finally, when $r \leq t$ and $s=r$ the trace similarly vanishes as the metrology qubits are completely unaffected. Define $s_\text{max}=r-1$ if $r \leq t$ and $s_\text{max}=t$ otherwise. Using these simplifications, we obtain
\begin{equation}
     \frac{1}{|\mathcal{K}|} \sum_{k \mathcal{K}} \Tr \big( \Pi_k \rho_\text{out}(k,\Gamma) \big) = \sum_{\alpha} \sum_{r=1}^m \sum_{P \in \mathcal{P}_m^{(r)}} \sum_{s=0}^{s_\text{max}} 3^{-s}|a_{\alpha,P}|^2 \frac{\binom{m-r}{t-s}}{\binom{m}{t}} \leq \sum_{r=1}^m \sum_{s=0}^{s_\text{max}} 3^{-s} \frac{\binom{m-r}{t-s}}{\binom{m}{t}},
\end{equation}
where the inequality follows from the completeness relationship, Eq.~\eqref{eqn:complete}. Re-arranging the above sum
\begin{equation}
\begin{split}
    \frac{1}{|\mathcal{K}|} \sum_{k \mathcal{K}} \Tr \big( \Pi_k \rho_\text{out}(k,\Gamma) \big) &\leq \frac{1}{\binom{m}{t}} \sum_{s=0}^t 3^{-s} \sum_{r=s+1}^{m} \frac{\binom{m-r}{t-s}}{\binom{m}{t}} \\
    &= \sum_{s=0}^t 3^{-s} \frac{\binom{m-s}{t-s+1}}{\binom{m}{t}} \\
    &= \frac{m-t}{t+1}\sum_{s=0}^t 3^{-s} \frac{(t+1)!(m-s)!}{(t-s+1)!m!} \\
    &= \frac{m-t}{t+1}+\frac{m-t}{t+1}\sum_{s=1}^t 3^{-s} \prod_{j=0}^{s-1} \frac{t+1-j}{m-j} \\
    &\leq \frac{m-t}{t+1}+\frac{m-t}{t+1}\sum_{s=1}^t \Big(\frac{t+1}{3m}\Big)^s \\
    &\leq \frac{3}{2} \frac{m-t}{t} \\
\end{split}
\end{equation}

\subsection{The Clifford Code}

For the Clifford code, we fix the location of the trap qubits, $\vec{l}$, to the final $t$ qubits. Hence $\Pi_k=\Pi$ is constant for all $k$ and after simplification due to the twirling lemma
\begin{equation}
\frac{1}{|\mathcal{K}|} \sum_{k \in \mathcal{K}} \rho_\text{out} (k,\Gamma) = \frac{1}{|\mathcal{K}|} \sum_{C \in \mathcal{C}_m} \sum_{\alpha} \sum_{P \in \mathcal{P}_m} |a_{\alpha,P}|^2 C^\dagger P C \rho_0 C^\dagger P C.
\end{equation}
Because we are summing over $\mathcal{C}_m$, we can greatly simplify the above, as for any $P \neq \mathbb{I}$
\begin{equation}
\frac{1}{|\mathcal{C}_m|} \sum_{C \in \mathcal{C}_m} C^\dagger P C \rho C^\dagger P C = \frac{1}{|\mathcal{P}_m|-1}\sum_{P^\prime \neq \mathbb{I} \in \mathcal{P}_m} P^\prime \rho P^\prime =\frac{4^m}{4^m-1} (\mathbb{I}/2^m-\rho/4^m)=\frac{1}{4^m-1} (2^m\mathbb{I}-\rho).
\end{equation}
Denoting $a=\sum_\alpha |a_{\alpha,\mathbb{I}}|^2$ we simplify the effective state to be
\begin{equation}
\frac{1}{|\mathcal{K}|} \sum_{k \in \mathcal{K}} \rho_\text{out} (k,\Gamma) = a \rho_0 +\frac{1-a}{4^m-1}(2^m\mathbb{I}-\rho_0)
\end{equation}
From which we compute
\begin{equation}
\frac{1}{|\mathcal{K}|}\sum_{k \in \mathcal{K}} \Tr \big( \Pi \rho_\text{out} (k, \Gamma) \big) = \Big(a-\frac{1-a}{4^m-1}\Big) \Tr \big( \Pi \rho_0 \big)+2^m\frac{1-a}{4^m-1} \Tr \big( \Pi \big).
\end{equation}
The first trace is null, since the ideal outcome is of course orthogonal to its own orthogonal compliment. The second trace computes as $\Tr(\Pi)=\Tr(\mathbb{I}-\rho) \text{Tr} (\dyad{0}^{\otimes t})=2^n-1$. The completeness relationship of the Kraus operators guarantees that $1-a \leq 1$, thus
\begin{equation}
\frac{1}{|\mathcal{K}|}\sum_{k \in \mathcal{K}} \Tr \big( \Pi \rho_\text{out} (k, \Gamma) \big) \leq  \frac{2^m(2^{m-t}-1)}{4^m-1} \leq \frac{2^m \cdot 2^{m-t}}{4^m} \leq 2^{-t}.
\end{equation}

\section{Appendix C: Extension of the Protocols to Two Uses of the Quantum Channel}

\setcounter{equation}{0}
\renewcommand\theequation{C.\arabic{equation}}

\begin{figure}[h]
\includegraphics[width=0.55\textwidth]{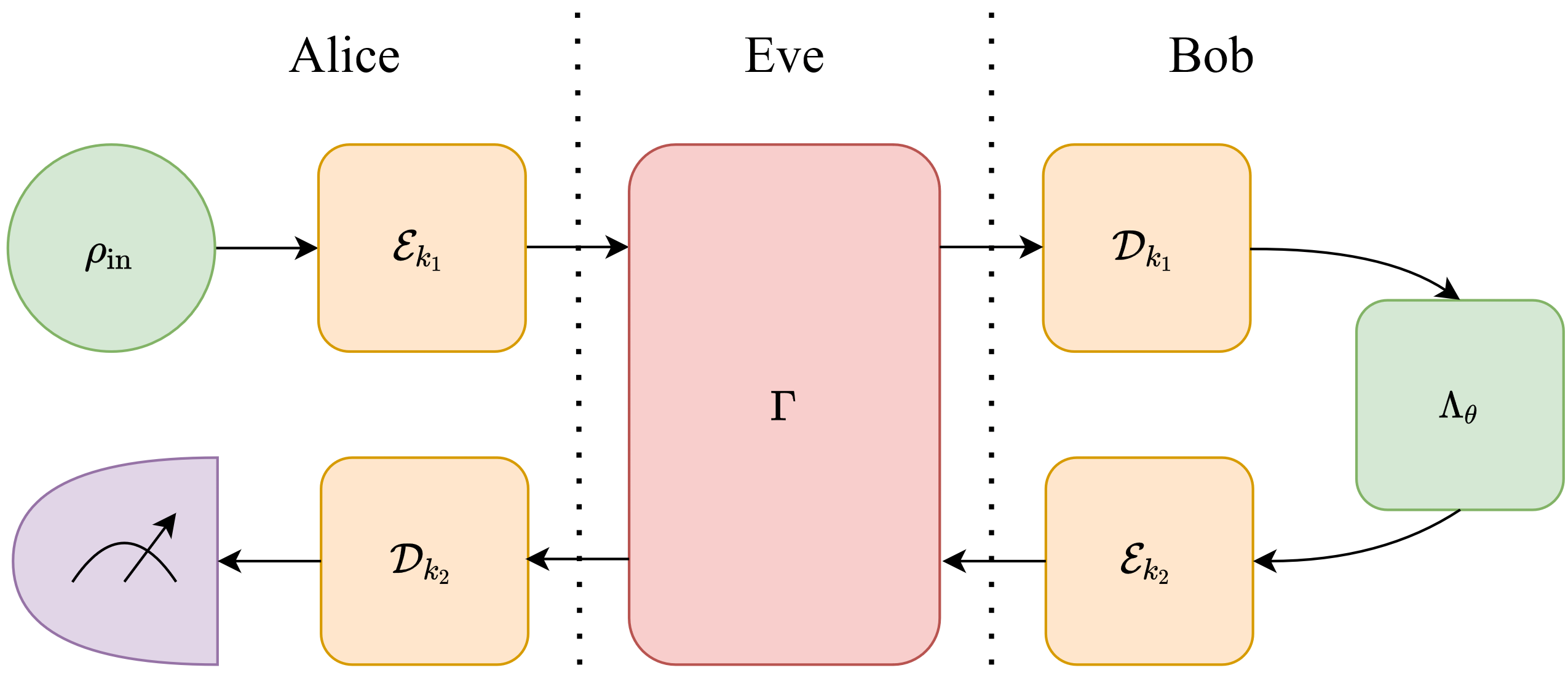}
\caption{In the extended version of the protocol, in which the quantum channel is used twice, Alice sends the quantum state $\rho_\text{in}$ to Bob to encode the parameter $\theta$. Here, $\rho_\text{in}$ a combination of the flag qubits and the quantum state which is intended to be encoded. Because the quantum channel is used twice, the classical key shared by Alice and Bob describes the encryption and decryption operation for the first use of the quantum channel ($\mathcal{E}_{k_1}, \mathcal{D}_{k_1}$) and the second use of the quantum channel ($\mathcal{E}_{k_2}, \mathcal{D}_{k_2}$). We again assume that Eve can perform any CPTP map $\Gamma$ when interacting with the channel.}
\label{fig:QASMetrology2}
\end{figure}

Both of the protocols can easily be adapted to a scenario where Alice solely delegates the task of parameter encoding to Bob. The extended protocol is illustrated in Figure~\ref{fig:QASMetrology2}. The primary difference is that the input state $\rho_\text{in}$ has not yet been encoded. Instead, the encoding is done by Bob upon receipt. In this extended scenario, it is paramount for Bob to be able to perform encryption and decryption operations, otherwise Eve could tamper with the quantum state in a completely undetectable fashion. We assume that the encoded quantum state $\rho_\theta=\Lambda_\theta(\rho)$ is a pure state.

We use similar mathematical tools as in the single use of the channel to determine bounds on the soundness.

\subsection{The Trap Code}

In the double use of the quantum channel, the final quantum state can be expressed as
\begin{equation}
    \rho_\text{out} (k,\Gamma) = C_2^\dagger \Gamma_b \bigg( C_2 \tilde{\Lambda}_\theta \Big( C_1^\dagger \Gamma_a \big( C_1 \pi \rho_0 \pi^\dagger C_1^\dagger \big) C_1 \Big) C_2^\dagger \bigg) C_2.
\end{equation}
Where we have divided $\Gamma$ into two CPTP maps $\Gamma_a$ and $\Gamma_b$, representing a malicious eavesdropper's first and second interaction with the quantum channel respectively. The CPTP map $\tilde{\Lambda}_\theta$ represents the parameter encoding performed by Bob on all but the $t$ flag qubits. The projector can be similarly expressed as $\Pi_k=\pi \Pi \pi^\dagger$, with $\Pi=(\mathbb{I}-\rho_\theta) \otimes \dyad{0}^{\otimes t}$.

To compute a bound on the soundness, we again decompose the CPTP maps $\Gamma_a$ and $\Gamma_b$ as sums of Kraus operators $\{ A_\alpha \}$ and $\{ B_\beta \}$, which are further decomposed into a sum of Pauli operations $P$ and $Q$. After, applying the twirling lemma, the final state can be written as \begin{equation}
\frac{1}{|\mathcal{K}|} \sum_{k \in \mathcal{K}} \pi^\dagger \rho_\text{out} (k,\Gamma) \pi = \frac{1}{|\mathcal{K}|} \sum_{C_1,C_2 \in \mathcal{C}_1^{\otimes m}} \sum_{\pi} \sum_{\alpha,\beta} \sum_{P,Q \in \mathcal{P}_m}  |a_{\alpha,P}|^2  |b_{\beta,Q}|^2 C_2^\dagger Q C_2 \tilde{\Lambda}_\theta \big( C_1^\dagger P C_1 \pi \rho_0 \pi^\dagger C_1^\dagger P C_1 \big) C_2^\dagger Q C_2.
\end{equation}
Next, we again use the fact that the Clifford group $\mathcal{C}_1$ will map $P \in \{X,Y,Z\}$ to an equal distribution over $\{ \pm X, \pm Y , \pm Z \}$. In this proof, we define $r$ to be the number of non-identity indices spanned by $P$ or $Q$. For example the total number of non-identity indices spanned by $P=\mathbb{I} \otimes X \otimes Z$ and $Q=\mathbb{I} \otimes X \otimes \mathbb{I}$ is $r=2$. Again, for any $s \leq r$, there are $\binom{m-r}{t-s}$ permutations $\pi$ of the flag qubits where the non-identity indices spanned by $C_1PC_1^\dagger$ or $C_2QC_2^\dagger$ interact with $s$ flag qubits. Note that $\tilde{\Lambda}_\theta$ does not interact with the flag qubits. The number of $C_1,C_2$ which results in an accepted outcome is less than $(\frac{5}{9})^s|C_1|^{2m}$. To understand why, suppose that $s=1$, then there are two possibilities. The first is that either $P$ or $Q$ has a single non-identity term, which after being mapped by the respective Pauli, the only accepted possibility is when said term is mapped to $\pm Z$, which occurs with a frequency of $1/3 < 5/9$. The second possibility is that $P$ and $Q$ have a non-identity term at the same index, now the only accepted terms (up-to a phase) is when said terms are mapped to one of the tuples $(X,X),(Y,Y),(Z,Z),(X,Y),(Y,X)$, which occurs with frequency $5/9$. The frequency of acceptance is multiplicative for each of the $s$ non-identity indices and thus the total number of Clifford operations which map $P,Q$ to an accepted output state is bounded by $(\frac{5}{9})^s|C_1|^{2m}$. Once again we define $s_\text{max}=r-1$ for $r \leq t$. Here we define $c_r$ to be the sum of all $|a_{\alpha,P}|^2 |b_{\beta,Q}|^2$ with $r$ total non-identity indices spanned by $P$ and $Q$. Combining everything we obtain
\begin{equation}
\begin{split}
\frac{1}{|\mathcal{K}|}\sum_{k \in \mathcal{K}} \Tr \big( \Pi_k \rho_\text{out} (k, \Gamma) \big) &\leq \sum_{r=0}^m c_r \sum_{s=0}^{s_\text{max}} \Big( \frac{5}{9} \Big)^s \frac{\binom{m-r}{t-s}}{\binom{m}{t}} \\
&\leq  \sum_{s=0}^t \Big( \frac{5}{9} \Big)^s \sum_{r=s+1}^{m} \frac{\binom{m-r}{t-s}}{\binom{m}{t}} \\
&= \sum_{s=0}^t \Big( \frac{5}{9} \Big)^s  \frac{\binom{m-s}{t-s+1}}{\binom{m}{t}} \\
&= \frac{m-t}{t+1}\sum_{s=0}^t \Big( \frac{5}{9} \Big)^s \frac{(t+1)!(m-s)!}{(t-s+1)!m!} \\
&= \frac{m-t}{t+1}+\frac{m-t}{t+1}\sum_{s=1}^t \Big( \frac{5}{9} \Big)^s \prod_{j=0}^{s-1} \frac{t+1-j}{m-j} \\
&\leq \frac{m-t}{t+1}+\frac{m-t}{t+1}\sum_{s=1}^t \Big( \frac{5}{9} \Big)^s \big(\frac{t+1}{m}\big)^s \\
&\leq \frac{9}{4}  \frac{m-t}{t+1}. \\
\end{split}
\end{equation}

\subsection{The Clifford Code}

Similar to trap code, we use the same formulation as we did for the original version of the protocol. After simplification, the expected state can be written as
\begin{equation}
\frac{1}{|\mathcal{K}|} \sum_{k \in \mathcal{K}} \rho_\text{out} (k,\Gamma) = \frac{1}{|\mathcal{K}|} \sum_{C_1,C_2 \in \mathcal{C}_m} \sum_{\alpha,\beta} \sum_{P,Q}  |a_{\alpha,P}|^2  |b_{\beta,Q}|^2 C_2^\dagger Q C_2 \tilde{\Lambda}_\theta \big( C_1^\dagger P C_1 \rho_0 C_1^\dagger P C_1 \big) C_2^\dagger Q C_2.
\end{equation}
Because we are summing over the complete Clifford group $\mathcal{C}_m$, we can simplify the above to
\begin{equation}
\begin{split}
& \frac{1}{|\mathcal{K}|} \sum_{k \in \mathcal{K}} \rho_\text{out} (k,\Gamma) \\
= & ab \tilde{\Lambda}_\theta \big(\rho_0 \big)+\frac{(1-a)b}{4^m-1}\tilde{\Lambda}_\theta \big(2^m\mathbb{I}-\rho_0 \big)+\frac{a(1-b)}{4^m-1} \Big(2^m\mathbb{I}-\tilde{\Lambda}_\theta \big(\rho_0 \big) \Big) + \frac{(1-a)(1-b)}{(4^m-1)^2} \tilde{\Lambda}_\theta \big( \rho_0 \big) \\
= & \Big(ab-\frac{a(1-b)+b(1-a)}{4^m-1}+\frac{(1-a)(1-b)}{(4^m-1)^2} \Big) \tilde{\Lambda}_\theta \big(\rho_0 \big)+\frac{(1-a)b+a(1-b)}{4^m-1} 2^m \mathbb{I}, \\
\end{split}
\end{equation}
where $a=\sum_\alpha |a_{\alpha,\mathbb{I}}|^2 \leq 1$ and $b=\sum_\beta |b_{\beta,\mathbb{I}}|^2 \leq 1$. From which we compute
\begin{equation}
\frac{1}{|\mathcal{K}|}\sum_{k \in \mathcal{K}} \Tr \big( \Pi \rho_\text{out} (k, \Gamma) \big) = 2^m\frac{(1-a)b+a(1-b)}{4^m-1} (2^{m-t}-1) \leq \big( (1-a)b+a(1-b) \big) 2^{-t} \leq 2^{-t}.
\end{equation}

\section{Appendix D: Undetectable Attack on the Protocol Described by Huang et al.}

\setcounter{equation}{0}
\renewcommand\theequation{D.\arabic{equation}}

The protocol described in \cite{huang2019} is supposed to function on the basis that the actions of Alice and Charlie and probabilistic. Every round, Alice sends one of four possible $n$ qubit quantum states through the channel
\renewcommand{\arraystretch}{1.5}
\begin{center}
\begin{tabular}{ c | c }
Input State & \hspace{2.5pt}Probability\hspace{2.5pt} \\
\hline
\hspace{2.5pt}$\ket{\psi_{+1}}=(\ket{0}^{\otimes n}+\ket{1}^{\otimes n})/\sqrt{2}$\hspace{2.5pt} & $P_A/2$ \\
\hspace{2.5pt}$\ket{\psi_{-1}}=(\ket{0}^{\otimes n}-\ket{1}^{\otimes n})/\sqrt{2}$\hspace{2.5pt} & $P_A/2$ \\
$\ket{d_0}=\ket{0}^{\otimes n}$ & $(1-P_A)/2$ \\
$\ket{d_1}=\ket{1}^{\otimes n}$ & $(1-P_A)/2$ \\
\end{tabular}
\end{center}
whereas Charlie applies either the unitary $U_{\theta+m\pi/n}^{\otimes n}$ or $U_{m\pi/n}^{\otimes n}$ with probabilities $P_C$ and $1-P_C$ respectively, where $U_x = e^{-i \frac{x}{2}Z}$ and $0 \leq m \leq n-1$ is a random integer.

If Alice sends one of the decoy states $\ket{d_{0/1}}$, then regardless of the unitary Charlie applies, the final state (up to a global phase) will be equal to the input state. This can be verified deterministically by measuring in the computational basis. If instead Alice sends a phase sensitive states, $\ket{\psi_{\pm 1}}$, Alice must communicate with Charlie to ask whether or not the unknown parameter $\theta$ was encoded. If no phase was encoded, then the final state (up to a global phase) is expected to be $\ket{\psi_{\pm (-1)^m}}$, which can be verified by measuring the state in the $X$ basis, as the resulting measurement will always be one of the $\pm(-1)^m$ eigenvalues of $X^{\otimes n}$.

This protocol argues security due to the large number of deterministic measurements in two non-commuting basis'. Therefore, if a malicious eavesdropper, Eve, tampers with the quantum channel, there is a high probability of detecting Eve after just a few rounds. In \cite{huang2019}, the authors describe an attack from Eve to minimize the probability of them being detected while gathering as much information about the unknown parameter as possible, and the authors claim this is done by Eve performing state discrimination. Using the attack described, if Eve tampers with $k$ states, the probability of remaining undetected is
$(1-\frac{1-P_AP_C}{4})^k$.

However, consider the following attack. Eve intercepts the input state and sets in aside. Eve then sends their own quantum state $\ket{\psi_{+1}}_E$ to Charlie. Charlie will then probabilistically apply a unitary to Eve's state and sends it back through the quantum channel. Eve then measures their possibly encoded quantum state in the $X$ basis. If a $+1$ eigenvalue is observed, Eve returns the intercepted state back to Alice, unmodified. If a $-1$ eigenvalue is observe, Eve first applies $Z$ on a qubit of the intercepted state to flip the relative phase, and then sends it back to Alice.

The above attack is completely undetectable by Alice. Whenever Alice sends a decoy state $\ket{d_{0/1}}$, then the final state will still be the initial state up to a global phase. The remaining deterministic outcomes are when Alice sends a phase sensitive state $\ket{\psi_{\pm 1}}$, and Charlie applies the unitary $U_{m\pi/n}^{\otimes n}$. In this instance Eve will deterministically observe a measurement outcome with a $(-1)^m$ eigenvalue, therefore we can write that the quantum state Eve sends to Alice is $Z^m\ket{\psi_{\pm 1}}=\ket{\psi_{\pm (-1)^m}}$; which is identical to the state Alice expects to receive.

In addition to being undetectable, Eve can obtain an estimate of the unknown parameter for themselves; the precision of which is ultimately determined by how much information they have about the values of $P_C$ and $m$. Because of the probabilistic action taken by Charlie, the state Eve has prior to measurement is
\begin{equation}
\begin{split}
\rho_E&=P_C U_{\theta+m\pi/n}^{\otimes n} \dyad{\psi_{+1}} U_{\theta+m\pi/n}^{\dagger \otimes n} +(1-P_C) U_{m\pi/n}^{\otimes n} \dyad{\psi_{+1}} U_{m\pi/n}^{\dagger \otimes n}  \\
&=P_C U_{\theta}^{\otimes n} \dyad{\psi_{(-1)^m}} U_{\theta}^{\dagger \otimes n} +(1-P_C) \dyad{\psi_{(-1)^m}}, \\
\end{split}
\end{equation}
and the expected value of the observable $O=X^{\otimes n}$ is
\begin{equation}
\begin{split}
\expval{O}_{\rho_E} &= P_C \Tr \big(X^{\otimes n} U_{\theta}^{\otimes n} \dyad{\psi_{(-1)^m}} U_{\theta}^{\dagger \otimes n} \big) +(1-P_C) \Tr \big(X^{\otimes n} \dyad{\psi_{(-1)^m}} \big) \\
&=(-1)^m P_C \cos(n\theta)+(1-P_C)(-1)^m \\
&=-2(-1)^m P_C \sin^2 \frac{n \theta}{2}+(-1)^m. \\
\end{split}
\end{equation}
To re-iterate, a precise estimate can only be made if Eve has some information about the value of $P_C$ and $m$. In the scenario which Eve knows the exact values chosen by Charlie and Alice, then Eve can achieve an estimate with a precision of
\begin{equation}
\Delta^2\theta_E = \frac{\Delta^2 O_{\rho_E}}{\nu_A |\partial \expval{O}_{\rho_E}|^2} = \frac{1+(P_C^{-1}-1) \sec^2 \frac{n \theta}{2}}{\nu_A n^2},
\end{equation}
where $\nu_A$ is the number of states Alice sends to Charlie through the quantum channel. Notice that the Heisenberg limit is recovered when $P_C=1$, or in the scenario when Charlie encodes the phase every round.

\end{document}